\documentclass{article}
\usepackage[margin=1.3in]{geometry}
\usepackage{authblk}
\usepackage[utf8]{inputenc}
\usepackage{amsmath}
\usepackage{amsfonts}
\usepackage{amsthm}
\usepackage{mathtools}
\usepackage{mathrsfs}
\usepackage{enumitem}
\usepackage{graphicx}
\usepackage{scrextend}
\usepackage{blindtext}
\usepackage{caption}
\usepackage{url}
\usepackage{subcaption}
\usepackage{circuitikz}
\usepackage{listings}
\usepackage{color}
 \usepackage{microtype}
\usepackage{graphicx}
\usepackage{bbm}
\usepackage{parskip}
\usepackage{bm}
\usepackage{booktabs}
\usepackage{array,multirow}
\usepackage{comment}
\newcolumntype{P}[1]{>{\centering\arraybackslash}p{#1}}
\usepackage{hyperref}
\usepackage[linesnumbered,lined,commentsnumbered,ruled]{algorithm2e}
\usepackage{amssymb}
\usepackage{pifont}
\newcommand{\cmark}{\ding{51}}%
\newcommand{\xmark}{\ding{55}}%
\let\oldReturn\Return
\renewcommand{\Return}{\State\oldReturn}
\newtheorem{theorem}{Theorem}[section]

\newtheorem{remark}[theorem]{Remark}

\title{\textbf{Testing exchangeability by pairwise betting}}

\author[1]{Aytijhya Saha}
\author[2]{Aaditya Ramdas}

\affil[1]{Indian Statistical Institute, Kolkata, India. Email: \href{mailto:aytijhyasaha02@gmail.com}{aytijhyasaha02@gmail.com}}
\affil[2]{Carnegie Mellon University, Pittsburgh, USA. Email: \href{mailto:aramdas@cmu.edu}{aramdas@cmu.edu}}
\begin{document}
\date{}

\maketitle

\begin{abstract}
 In this paper, we address the problem of testing exchangeability of a sequence of random variables, $X_1, X_2,\cdots$. 
 This problem has been studied under the recently popular framework of \emph{testing by betting}. But the mapping of testing problems to game is not one to one: many games can be designed for the same test. Past work established that it is futile to play single game betting on every observation: test martingales in the data filtration are powerless. Two avenues have been explored to circumvent this impossibility: betting in a reduced filtration (wealth is a test martingale in a coarsened filtration), or playing many games in parallel (wealth is an e-process in the data filtration). The former has proved to be difficult to theoretically analyze, while the latter only works for binary or discrete observation spaces.
 Here, we introduce a different approach that circumvents both drawbacks. We design a new (yet simple) game in which we observe the data sequence in pairs. Despite the fact that betting on individual observations is futile, we show that betting on pairs of observations is not.  
 To elaborate, we prove that our game leads to a nontrivial test martingale, which is interesting because it has been obtained by shrinking the filtration very slightly.
 We show that our test controls type-1 error despite continuous monitoring, and achieves power one for both binary and continuous observations, under a broad class of alternatives. Due to the shrunk filtration, optional stopping is only allowed at even stopping times, not at odd ones: a relatively minor price. We provide a wide array of simulations that align with our theoretical findings.
\end{abstract}

\section{INTRODUCTION}
 A sequence of random variables  $\{X_t\}_{t\geq 1}$ is exchangeable if and only if for every $t$ and every permutation $\sigma$ of the first $t$ indices, the joint distribution of $(X_1,\cdots, X_t)$ is  same as the joint distribution of
$(X_{\sigma(1)},\cdots, X_{\sigma(t)})$. Suppose that we sequentially observe a series of random variables $X_1, X_2, \cdots$ one by one. Consider the fundamental problem of testing if our data form an exchangeable sequence: 
$$H_0: X_1,X_2,\cdots \text{ are exchangeable.}$$

From a single sequence of data, one cannot distinguish whether the data is iid (independent and identically distributed) or exchangeable~\cite{ramdas2022testing}, and so one can equivalently view this paper as designing a test for the iid assumption: 
\[
H'_0: X_1,X_2,\cdots \text{ are iid.}
\]
We design a new sequential test for $H_0$ or $H_0'$, establish its consistency in both binary and continuous settings, primarily focusing on first-order Markov and AR(1) alternatives respectively, and prove that its ``growth rate'' asymptotically matches that of an oracle that knows the ground truth. Using these as building blocks, we then show that our test is also consistent against a much more general class of alternatives.


The vast majority of theoretical results in machine learning heavily rely on the exchangeability assumption, or the stronger iid assumption. 
These methods may encounter significant challenges when this assumption is violated (for example due to Markovian dependence between the data), underscoring the importance of rigorously testing data for its exchangeability.  

The current paper seeks to add a new method of testing exchangeability based on \emph{pairwise betting} to the two (recent) methods that are known so far, which are based on conformal prediction~\cite{vovk2021testing}  and universal inference~\cite{ramdas2022testing}. The former has proven difficult to analyze and there is currently no theoretical guarantee of consistency against any class of alternatives, while the latter is applicable only for a sequence of binary observations (or a small discrete alphabet). What distinguishes our approach is its applicability to general observation spaces (like the former approach), while we establish theoretical guarantees like consistency in both binary and continuous cases against a broad class of alternatives.


The overarching technical umbrella that ties together all three of the above solutions is that they stem from constructing ``test martingales'' (or, more generally, ``e-processes''), which can be interpreted as wealth of a gambler playing a stochastic game (or, for e-processes, many games in parallel)~\cite{ramdas2022game}. Going further with such a game-theoretic language, all 3 methods can be thought of as instances of the principle of ``testing by betting'', which we summarize below. 

\paragraph{Testing by betting.} In order to test a hypothesis, one designs a game of chance which has two properties. If the null hypothesis is true, the game rules must be such that no gambler (starting with one dollar) can systematically make money, meaning that their wealth is a nonnegative martingale (defined later): nonnegative because they cannot bet more than they have, and martingale because the wealth remains constant in expectation and thus it is unlikely that they will ever multiply their wealth by a large amount (in other words, they're playing roulette). More generally, the wealth is allowed to be a nonnegative \emph{super}martingale under the null, which means it can decrease in expectation. However, if the null hypothesis is false, the game rules must allow smart gamblers to multiply their wealth exponentially. The achieved exponent,  or the average expected logarithm of the wealth (defined formally soon), is defined as the ``rate of growth'' of wealth. The optimal betting strategy is one which maximizes the rate of growth of wealth under the alternative.

For testing a simple null hypothesis (that is, the null hypothesis is a single distribution) against a simple alternative, testing by betting is very straightforward. The optimal bet is simply given by a likelihood ratio of the alternative to the null~\cite{shafer2021testing}. However, for composite null hypotheses like our $H_0$, and composite alternatives like $H_1$, there is no unique translation of the testing problem to a game. There are many games that one could write down that test such nulls, and each of them permits different betting strategies with different rates of growth of wealth. It appears to be difficult to apriori determine which games' optimal wealth is the ``best'' one. 

A further nuance is the following key fact: for some nulls like our $H_0$, a single game in which one observes a single data point in each round, provably does not suffice to test the null because the constraints imposed by the game are too strong. \cite{ramdas2022testing} prove that every nonnegative (super)martingale in such a game is constant or decreasing (zero rate of growth). 

There are two options to circumvent the aforementioned negative result: (i) one must work in a reduced/coarsened filtration, which amounts to throwing some information away or restricting how much information is released, or (ii) one must play many games in parallel, each one against a different subset of the null hypothesis, and be judged on the minimum wealth across all games; this minimum wealth is no longer a nonnegative supermartingale, and is called an ``e-process'' (defined formally below). It turns out that the conformal strategy~\cite{vovk2021testing} takes route (i), while the universal inference strategy~\cite{ramdas2022testing} takes route (ii), and this distinction is discussed further in the latter paper (and a bit more in the current paper).

The main novelty in our work is the consideration of a different betting game under route (i), which is applicable to a general observation space. The wealth process produced is indeed a nonnegative martingale, except in a (very slightly) reduced filtration, as explained later. Our solution processes data in pairs, and we will construct an exact e-value in each round of the game. E-values are the building block for nonnegative martingales:  our wealth process is simply a product of these e-values, and will be a nonnegative martingale under $H_0$. In recent years, e-values have emerged as a promising alternative to p-values for handling such problems (\cite{grunwald2020safe,vovk2021values,ramdas2022game}). Below, we provide a concise technical overview of the key concepts and essentials in this rapidly evolving field.

\paragraph{E-process.}
Consider a nonnegative sequence of adapted random variables $E \equiv \{E_t\}_{t\geq0}$ and
let $H_0$, the null hypothesis, be a set of distributions. We call $E$ as an e-process for $H_0$, if 
\begin{equation}
    \mathbb{E}_{\mathbb{P}}[E_\tau] \leq 1, \text{ for all stopping times }\tau, \text{ for all } \mathbb{P}\in H_0
\end{equation}
Large values of the e-process encode evidence against the null. (Ideally, the evidence $E$ should increase to infinity under $H_1$, almost surely.) Further, suppose we stop and reject the null at the stopping time
\begin{equation}\label{eq:stopping}
    \tau_{\alpha}=\inf\left\{t\geq 1 : E_t\geq \frac{1}{\alpha}\right\}.
\end{equation}
This rule results in a level $\alpha$ sequential test, meaning that if the null is true, the probability that it ever stop falsely rejects the null is at most $\alpha$. This is easily seen by applying Markov’s inequality to the stopped e-process $E_{\tau_{\alpha}}$ (or, equivalently, Ville's inequality \cite[Lemma 1]{howard2020time}). 

\paragraph{Test Martingale.} 
An integrable process $M \equiv \{M_t\}$ that is adapted to a filtration $\mathscr{F} \equiv \{\mathscr{F}_t\}_{t\geq 0}$,  is called a \emph{martingale} for $\mathbb P$ with respect to filtration $\mathscr{F}$  if 
\begin{equation}\label{eq:martingale}
   \mathbb{E}_{\mathbb{P}}[M_t \mid \mathscr{F}_{t-1}] = M_{t-1} 
\end{equation} for all $t\geq 1$. 
$M$ is called a \emph{test martingale} for $H_0$ if it is a martingale for every $\mathbb P \in H_0$, and if it is non-negative with $M_0=1$.
Game-theoretically, a test martingale for $\mathbb{P}$ is the wealth process of a gambler who sequentially bets against $H_0$, starting with an initial wealth of $1$. The optional stopping theorem implies that for \emph{any} stopping time $\tau$ and any $\mathbb{P} \in H_0$, we have $\mathbb{E}_{\mathbb{P}}[M_\tau] \leq 1$. Thus, if $M$ is a test martingale for $H_0$, it is also an e-process for $H_0$. 

An e-process (or test martingale) is called consistent for $H_1$, if $\lim_{t\to \infty} M_t = \infty$ almost surely for any $\mathbb P \in H_1$, meaning that under the alternative, it accumulates infinite evidence against the null in the limit. Usually, the evidence grows exponentially, so the ``growth rate'' of $M$ is defined as $\inf_{\mathbb P \in H_1} \lim_{t \to \infty} \mathbb E_{\mathbb{P}}[\log M_t]/t$. A positive growth rate implies consistency.

\paragraph{Betting score.} We call the factor by which the gambler multiplies the money he risks at $t$-th round of betting as the \emph{betting score} $B_t$. $B_t$ is an e-value, meaning that it has expectation at most 1 under $H_0$.
Note that, $M_t=M_{t-1}\times B_t=\prod_{i=1}^tB_t$. 
Hence, betting scores can be viewed as building blocks of test martingales.

\begin{table}[!htb]
    \centering
    \caption{Comparing our pairwise betting  with the universal inference e-process \cite{ramdas2022testing} and the conformal test martingale \cite{vovk2021testing}.}
\label{tab:comp}
    \resizebox{\linewidth}{!}{
\begin{tabular}{ |P{4cm}|P{4cm}|P{4cm}|P{4cm}|  }
 \hline
 & Applicable to general  observation spaces & Optional stopping in the data filtration & Provably Consistent with growth rate analysis \\
 \hline
 Pairwise betting   & \cmark  & \xmark & \cmark \\
  \hline
 Universal inference &  \xmark   & \cmark  & \cmark \\
  \hline
 Conformal inference & \cmark & \xmark & \xmark \\
 \hline
\end{tabular}}
 \end{table}
 
\paragraph{Related Work.}
Sequential hypothesis testing has a long-standing history, beginning with the sequential probability ratio test of \cite{wald1945sequential}. However, while the basic theory holds for parametric hypotheses, it is often inadequate in the face of nonparametrically defined composite nulls and alternatives. More recently, the ``testing by betting" methodology \cite{shafer2019game,shafer2021testing}
has led to a “game-theoretic” approach to sequential hypothesis testing that has shown promise for nonparametric nulls \cite{shekhar2023nonparametric,waudby2023estimating,podkopaev2023sequential}.

One popular approach for testing exchangeability relies on conformal prediction \cite{vovk2003testing,fedorova2012plug,vovk2021testing,vovk2022conformal}. The core idea implicitly replaces the canonical filtration with a coarser filtration formed by the independent conformal p-values in each round. These are converted to e-values by ``calibration'', which are multiplied to form a test martingale for testing $H_0$. However, it is worth noting that the consistency (and growth rate) of conformal testing remains theoretically unproven.

 On the other hand, \cite{ramdas2022testing} approaches the problem of testing exchangeability by using universal inference \cite{wasserman2020universal} to derive an e-process for $H_0$, although this method is only suitable for binary (or small, discrete alphabet) data sequences. It is unclear how to extend it to more general observation spaces.

Our approach circumvents the limitations of these existing methods: it comes with provable growth rate guarantees unlike conformal testing, and it applies to any observation space unlike the universal e-process. Table \ref{tab:comp} summarizes the comparison.


\paragraph{Paper outline.}
The rest of the paper is organized as follows.
 In Section~\ref{sec:method}, we construct a pairwise betting game for testing exchangeability. More precisely, Subsection~\ref{test-binary} presents a test designed for binary data sequence and demonstrates its power against the natural class of first-order Markov alternatives. We also extend it to a larger class of alternatives. Then, in Subsection \ref{test-cont}, we develop a test for the continuous case. Section \ref{sec:exp} presents a comprehensive set of simulation studies that validate our theoretical findings. This article is concluded in Section \ref{conc}, following Section \ref{sec:discussion}, which provides a discussion on the key aspects of our approach. All proofs and relevant mathematical details are
provided in the Supplementary Materials.


\section{PAIRWISE BETTING}
\label{sec:method}

We begin here with the binary case for simplicity. The idea of pairwise betting readily extends from binary to any general observation space, offering versatility, and we get to this later.

Given a binary sequence of observations, it might seem intuitive to consider a betting game where a gambler places bets on individual data points in each round. However, as demonstrated by \cite{ramdas2022testing}, this game results in a powerless test. To overcome this challenge, we design a game that reveals data in pairs. 

In each odd step $t$, nature tells us the unordered set of the $t$-th and $(t+1)$-th observations. Based on this information (and all past observations), we bet on the order in which they are observed. It turns out that \emph{the composite null hypothesis  collapses to a point null, when we condition on the unordered pair}. 

For example, suppose that nature reveals the observations to be $\{0,1\}$, but we don't know whether the order was 01 or 10. Under the exchangeable or iid null, both of these are equally likely (and knowing previous observations gives no further information), so the null is simply a Bernoulli(0.5) for the two possibilities. However, under a Markovian alternative, one of them is more likely than the other (which one is more likely depends on the past, which we know), and we can use this information to bet. We bet by invoking the observations of \cite{shafer2021testing}, who proved that for a point (conditional) null, the optimal bet is the likelihood ratio of the (conditional) alternative to the null. 

Of course, when we start out, we don't know the alternative (the true Markov model, or its implied probabilities for the observed pair) and so our betting is noisy. A pragmatic strategy is to bet using a maximum likelihood estimate (regularized or smoothed, if needed) of the alternative based on the first $t-1$ observations which have been revealed to us. As we learn the true alternative, our betting becomes more accurate and we can provably make money, as argued formally below. This is known as the plug-in method~\cite{waudby2023estimating,ramdas2022game}, because we simply plug in empirical estimates of unknown parameters into our alternative model when we bet.

Algorithm \ref{algo:1} contains an overview of pairwise betting.

\begin{algorithm}
\SetKwData{Left}{left}\SetKwData{This}{this}\SetKwData{Up}{up}
\SetKwFunction{Union}{Union}\SetKwFunction{FindCompress}{FindCompress}
\SetKwInOut{Input}{Input}\SetKwInOut{Output}{output}
\SetAlgoLined
\Input{  Sequence of observations $x_1,x_2,\cdots$}
\Output{ Test Martingale $M_2,M_4,\cdots$}
$M_{2}=1$\;
\For{$t=3,5,\cdots$}{
\If{$x_{t}=x_{t+1}$}{$M_{t+1}=M_{t-1}$}

 \Else{$\mathbf{Z_{t}}:=(X_{t},X_{t+1})$\;
 $E_t:=$ Event that either $\mathbf{Z_{t}}=(x_{t},x_{t+1})$ or $\mathbf{Z_{t}}=(x_{t+1},x_{t})$\;
 $\mathbf{x^{t}}:=(x_1,\cdots,x_{t})$\;
 $L_t:=\mathbb{P}_{H_1}\left[\mathbf{Z_{t}}=(x_{t},x_{t+1})|\mathbf{\mathbf{x^{t-1}}},E_t\right]$\;
 
 Estimate $L_t$ by $\hat{L}_t$, based on $\mathbf{\mathbf{x^{t-1}}}$\;
 $\hat{B}_{t+1}:=2\hat{L}_t$\;
 $M_{t+1}=M_{t-1}\times \hat{B}_{t+1}$\;}}
\caption{Pairwise Betting}
\label{algo:1}
\end{algorithm}

Next, our objective is to formulate and analyze our test for two different scenarios: first, we concentrate on the binary case, with a focus on a first order Markov alternative, and second, we shift to the continuous case, with an emphasis on an AR(1) alternative.

\subsection{Test for Binary Observations}
\label{test-binary}

Suppose, we have a sequence of binary random variables $X_1, X_2, \cdots$. The realization of the random variables are denoted as $x_1,x_2,\cdots$. We primarily focus on first-order Markov alternative, i.e,
$\mathbb{P}[X_{t+1}|X_t,\cdots,X_1]=\mathbb{P}[X_{t+1}|X_t]$, for all $t\geq 1$. 

We consider a betting game, starting with an initial wealth $M_{0}=M_2=1$. Define, $\mathbf{Z_{t}} = (X_t, X_{t+1})$. At each odd time step $t$,
nature tells us the unordered set of the $t$ th and $(t+1)$ th observations. If it is either $(0,0)$ or $(1,1)$, no betting occurs in this case. Otherwise, we place bets on $\mathbf{Z_{t}}$, according to the likelihood ratio, conditioned on the observed values of $\mathbf{X^{t-1}}:=(X_1,\cdots,X_{t-1})$ and the event that either $\mathbf{Z_{t}}=(1,0)$ or $\mathbf{Z_{t}}=(0,1)$ (denote this event as $E_t$) and then nature unveils the observed value of $\mathbf{Z_{t}}$.
So, the conditional likelihood under $H_0$
is \begin{equation}
    \mathbb{P}_{H_0}(\mathbf{Z_{t}}=(x_t,x_{t+1})|\mathbf{X^{t-1}}=\mathbf{x^{t-1}},E_t)=\frac{1}{2},
\end{equation} since (1,0) and (0,1) are equally likely under $H_0 $. For $t=3,5,\cdots$, the conditional likelihood under $H_1$ is 
\begin{equation}
\label{alt-likelihood}
    \mathbb{P}_{H_1}(\mathbf{Z_{t}}=(x_t,x_{t+1})|\mathbf{X^{t-1}}=\mathbf{x^{t-1}},E_t)=\frac{{p}_{x_t|x_{t-1}}{p}_{x_{t+1}|x_{t}}}{{p}_{x_t|x_{t-1}}{p}_{x_{t+1}|x_{t}}+{p}_{x_{t+1}|x_{t-1}}{p}_{x_{t}|x_{t+1}}},
\end{equation}
where ${p }_{i|j}$ is the transition probability from $j$ to $i$ of the underlying Markov model.
Then, the betting score at $\frac{t+1}{2}$th
round of betting is the
 likelihood ratio of the (conditional) alternative (Equation \eqref{alt-likelihood}) to the null (which is $\frac{1}{2}$).
But, in a practical situation, we typically lack knowledge of the true transition probabilities. Therefore, it becomes necessary to estimate them. One viable option is to replace ${p}_{i|j}$ by its maximum likelihood estimator (MLE) based on the first $t-1$ many observations (denote it by $\hat{p}_{i|j}$) in Equation \eqref{alt-likelihood}. But, there could be other choices too (see Remark \ref{rmk:estimator}).
We do not bet in the first round. So, $\hat{B}_2=1$. And betting score at $\frac{t+1}{2}$th
round ($t=3,5,\cdots$) of betting is
\begin{equation}
\hat{B}_{t+1}=\mathbbm{1}_{E^c_t}+\frac{2\hat{p}_{X_t|X_{t-1}}\hat{p}_{X_{t+1}|X_{t}}}{\hat{p}_{X_t|X_{t-1}}\hat{p}_{X_{t+1}|X_{t}}+\hat{p}_{X_{t+1}|X_{t-1}}\hat{p}_{X_{t}|X_{t+1}}}\mathbbm{1}_{E_t}.
\end{equation}
Thus, the bettor’s wealth after $\frac{t+1}{2}$
rounds of betting is
\begin{equation}
\label{eq:test}
    M_{t+1}=M_{t-1}\times \hat{B}_{t+1}=\prod_{i=1}^{{(t+1)}/{2}} \hat{B}_{2i};~t=3,5,\cdots.
\end{equation}
It is easy to check that $\{M_2,M_4,\cdots\}$ is a test martingale for $H_0$. Hence, recalling~\eqref{eq:stopping},
\begin{equation}
   \tau^*=\inf\{t : M_{t}\geq 1/\alpha\} 
\end{equation}
 is the stopping time at which we reject the null, yielding a level $\alpha$ sequential test. It is worth noting that due to the shrunk filtration, optional stopping is only permissible at even stopping times, not at odd ones.

\subsubsection{Consistency of the test}
In this subsection, we present the main theorem characterizing the consistency and the rate of convergence of our test against first order Markov alternative.
Within the class of first-order Markov chains, the special case of iid Bernoulli data is characterized by the restriction $p_{1|0}=p_{1|1}$. Therefore, our test achieves power $1$ against the first-order Markov alternative, as established by the following theorem. Below, $j^c$ denotes the complement of $j$, i.e.\ $j^c=1$ when $j=0$, and $j^c=0$ when $j=1$.

\begin{theorem}
\label{thm:rate}
Under first order Markov alternative, assume that $p_{0|1},p_{1|0}\neq 0$ or $1$. Then,
 $\frac{\log (M_{2n})}{2n}\to r$ almost surely as $t\to \infty$, where
$$r=\frac{1}{2}\sum_{i=0}^1\sum_{j=0}^1 \log\left(\frac{2{p}_{j|i}{p}_{j^c|j}}{{p}_{j|i}{p}_{j^c|j}+{p}_{j^c|i}{p}_{j|j^c}}\right) p_ip_{j|i}p_{j^c|j}.$$ Further, $r = 0$ 
if $p_{1|0}=p_{1|1}$, and $r > 0$ otherwise. 
\end{theorem}
[See Section A.1 of Supplementary for the proof]
\begin{remark}
\label{rmk:estimator}
   It's important to note that the estimator of ${p}_{j|j}$, that we plugged in \eqref{alt-likelihood} 
 does not necessarily have to be the maximum likelihood estimator (MLE). The Theorem \ref{thm:rate} holds for any strongly consistent estimator, $\hat{p}_{j|j}$ of the transition probabilities, ${p}_{j|j}$. For instance, one can opt for the Bayesian maximum aposteriori (MAP) estimator, with a uniform prior, as an alternative to the MLE.
\end{remark}

Theorem \ref{thm:rate} ensures the consistency of our sequential level $\alpha$ test (defined in \eqref{eq:test}), by showing that our test martingale $\{M_{2n}\}_{n\geq1}$ increases to infinity exponentially fast in $n$, under first order Markov alternatives. Next, we extend and generalize this result.

\subsubsection{Generalization to a larger class of alternatives}

Although our primary focus is on first-order Markov alternatives, we show that our test martingale is consistent for much more general alternatives.
Consider any binary sequence for which the following almost sure limits exist:
\begin{subequations}
\label{eq:cond}
\begin{align}
   &\alpha:=\lim_{t\to \infty}\frac{n_{1|1}(t)}{t},\quad \beta:=\lim_{n\to \infty}\frac{n_{0|0}(t)}{t}\\
    &\gamma:=\lim_{t\to \infty} \frac{n_{1|0}(t)}{t}=\lim_{t\to \infty} \frac{n_{0|1}(t)}{t},\\
    &p_{i,j,j^c}:=\lim_{t\to \infty} \frac{n_{j^c|j|i}(t)}{t}, \text{ for } i,j\in(0,1).
\end{align}
\end{subequations}
where $n_{i|j}(t)$ denotes the number of $i$ following $j$ up to time $t$ and $n_{i|j|k}(t)$ represents the count of instances where $i$ follows $j$ following $k$ up to time $t$.
Define,
\begin{align*}
&p_{1|1}:=\frac{\alpha}{\alpha+\gamma}, ~p_{0|1}:=\frac{\gamma}{\alpha+\gamma},~p_{1|0}:=\frac{\gamma}{\beta+\gamma}, ~p_{1|1}:=\frac{\beta}{\beta+\gamma}. 
\end{align*}
Note that for first-order Markov, these parameters are nothing but the transition probabilities.
Let us also define the following constants: $$a=\frac{p_{1,1,0}}{p_{1,1,0}+p_{0,1,0}},~ b=\frac{p_{1,0,1}}{p_{1,0,1}+p_{0,0,1}}.$$
\begin{theorem}
\label{thm:gen}
For any binary data sequence, suppose that the limits $\alpha,\beta,\gamma$ and $p$ defined in \eqref{eq:cond} exist. Then,
 $\frac{\log(M_{2n})}{2n}\to r'$ almost surely as $n\to \infty$, where
$$r'=\frac{1}{2}\sum_{i=0}^1\sum_{j=0}^1 \log\left(\frac{2{p}_{j|i}{p}_{j^c|j}}{{p}_{j|i}{p}_{j^c|j}+{p}_{j^c|i}{p}_{j|j^c}}\right) p_{i,j,j^c}$$ is strictly greater than $0$ if $(2p_{0|1}-1)(a+p_{0|1}-1)\geq 0, (2p_{1|0}-1)(b+p_{1|0}-1)\geq 0$ and $p_{1|0}\neq p_{1|1}$.

\end{theorem}
[See Section A.2 of Supplementary for the proof]
\begin{remark}
    It is straightforward to verify that for a first-order Markov, $p_{i,j,j^c}=p_ip_{j|i}p_{j^c|j}$, which implies that  $r=r'$. Furthermore, for this particular case, it can be easily shown that $a=p_{0|1}$ and $b=p_{1|0}$ leading trivially to the inequalities $(2p_{0|1}-1)(a+p_{0|1}-1) = (2p_{0|1}-1)^2 \geq 0$ and $(2p_{1|0}-1)(b+p_{1|0}-1)=(2p_{1|0}-1)^2\geq 0$. Hence, Theorem \ref{thm:gen} can be regarded as a strict generalization of Theorem \ref{thm:rate}.
\end{remark}

\subsection{Test for the continuous case}
\label{test-cont}
To show the power and versatility of our test, we now extend it to a sequence of continuous random variables. 

Suppose, we observe $X_1, X_2, \cdots$ from some continuous distribution. It is worth noting that our test is versatile enough to handle a broad spectrum of scenarios. But as an illustrative example, we primarily focus on a stationary Gaussian AR(1) alternative, i.e,
\begin{align*}
    H_1 & : X \text{ is a stationary AR(1) process, with }  \\
    & X_{t+1}=aX_{t}+\varepsilon_{t+1},
    ~~t=1,2,\cdots, \text{ where } \\&\varepsilon_t\stackrel{i.i.d}{\sim} N(0,\sigma^2); a \text{ and } \sigma \text{ are unknown.}
\end{align*}
Using the same idea that we employed in the binary case, we start with an initial wealth $W_{0}=W_2=1$, and in each odd step $t$, nature tells us the unordered set of the $t$-th and $(t+1)$-th observations. (We don't bet in the first round because we have not seen any data.) In the continuous case, the probability that $X_t=X_{t+1}$ is zero. So, we now always place bets on $\mathbf{Z_{t}}$, for each odd $t$.
These bets are determined by the likelihood ratio, conditioned on the observed values of $\mathbf{X^{t-1}}:=\{X_1,\cdots,X_{t-1}\}$ and the event that either $\mathbf{Z_{t}}=(x_t, x_{t+1})$ or $\mathbf{Z_{t}}=(x_{t+1},x_t)$ (denote this event as $E_t$) and then nature unveils the observed value of $\mathbf{Z_{t}}$. Let us denote $E_t$ as the event that $\mathbf{Z_{t}}$ is either $(x_t, x_{t+1})$ or $(x_{t+1},x_t)$. So, the conditional likelihood under $H_0$
is 
\begin{equation}
    \mathbb{P}_{H_0}(\mathbf{Z_{t}}=(x_t,x_{t+1})|\mathbf{X^{t-1}}=\mathbf{x^{t-1}},E_t)=\frac{1}{2},
\end{equation}
 since $(x_t, x_{t+1})$ and $(x_{t+1},x_t)$ are equally likely under $H_0$.
It is easy to obtain that the conditional likelihood under the AR(1) alternative is
\begin{align}
\label{alt-likelihood-cont}
  \nonumber&\mathbb{P}_{H_1}(\mathbf{Z_{t}}=(x_t,x_{t+1})|\mathbf{X^{t-1}}=\mathbf{x^{t-1}},E_t)=\frac{f(x_{t-1},x_t,x_{t+1})}{f(x_{t-1},x_t,x_{t+1})+f(x_{t-1},x_{t+1},x_t)}, 
\end{align}
 $$ \text{ where }f(x,y,z):=\frac{1}{2\pi\sigma^2}e^{-\frac{1}{2\sigma^2}(y-ax)^2-\frac{1}{2\sigma^2}(z-ay)^2}.
 $$
Then, the betting score (for Oracle) at $\frac{t+1}{2}$th
round is 
\begin{equation}
    S_{t+1}=
  \frac{2f(X_{t-1},X_t,X_{t+1})}{f(X_{t-1},X_t,X_{t+1})+f(X_{t-1},X_{t+1},X_t)}.
\end{equation}
Recall that we don't bet at $t=1$, so $S_2=1$.
For practical use, we need to estimate $a$ and $\sigma^2$. One viable option is to 
estimate $f$ by $\hat{f}_{t-1}$, which is obtained by replacing the model parameter $a$ by its least squares estimator, $\hat{a}_{t-1}=\sum_{i=2}^{t-1}X_iX_{i-1}/\sum_{i=1}^{t-2}X_i^2$ and $\sigma^2$ by 
$\hat{\sigma}^2_{t-1}=\frac{1}{t-2}\sum_{i=2}^{t-1}(X_{i}-\hat{a}_{t-1}X_{i-1})^2,$ for $t=3,5,\cdots$.
So, the bet at $\frac{t+1}{2}$th round ($t=3,5,\cdots$) is
\begin{equation}
\hat{S}_{t+1}=\frac{2\hat{f}_{t-1}(X_{t-1},X_t,X_{t+1})}{\hat{f}_{t-1}(X_{t-1},X_t,X_{t+1})+\hat{f}_{t-1}(X_{t-1},X_{t+1},X_t)}.
\end{equation}
Thus, the bettor’s wealth after $\frac{t+1}{2}$
rounds of betting is
\begin{equation}
\label{eq:test-martingale-cont}
    W_{t+1}=W_{t-1}\times \hat{S}_{t+1}=\prod_{i=2}^{{(t+1)}/{2}} \hat{S}_{2i}.
\end{equation}
It is easy to check that $\{W_2,W_4,\cdots\}$ is a test martingale, (with respect to a shrunk filtration) for testing exchangeability. Recalling~\eqref{eq:stopping}, 
\begin{equation}
\label{eq:test-cont}
   \tau'=\inf\{t : W_{t}\geq 1/\alpha\} 
\end{equation}
 is a level $\alpha$ sequential test. As before, due to the shrunk filtration, optional stopping is only permissible at even stopping times, not at odd ones.
 
 \subsubsection{Consistency of the test}

In this subsection, we present a crucial result concerning the consistency of our test with AR(1) alternative.
It's worth noting that, within the class of AR(1), the special case of iid $N(0,\sigma^2)$ is characterized by the restriction $a=0$. Therefore, our test is consistent against the AR(1) alternative, as established by the following theorem.
\begin{theorem}
\label{thm:ar-rate}
    Let, $\{X_t\}_t$ be a stationary Gaussian AR(1) process. Then, $$\frac{\log(W_{2n})}{2n}\stackrel{a.s}{\longrightarrow}r^*, \text{ as } n\to\infty,$$ where 
    $r^*=\frac{1}{2}\mathbb{E}\left(\log\left(S_4\right)\right)\geq 0$ and equality holds if and only if $a=0$ (in which case the null would be true).
\end{theorem}
[See Section A.3 of Supplementary for the proof]

In essence, Theorem \ref{thm:ar-rate} ensures the consistency of our sequential level $\alpha$ test (defined in \eqref{eq:test-cont}), by revealing that our test martingale $\{W_{2n}\}_{n\geq1}$ increases to infinity exponentially fast in $n$, under AR(1) alternative. Although it appears to be difficult to find a closed-form expression of $r^*$ in terms of the model parameters, one can get an approximation (for example, using Monte Carlo simulation).
Next, we extend and generalize this result.

\subsubsection{Generalization to a larger class of alternatives}
Although our primary focus is on AR(1) alternatives, we show that our test is consistent for much more general alternatives.

\begin{theorem}
\label{thm:cont-gen}
    Let $\{X_t\}_{t \geq 1}$ be an ergodic process. Then, $$\frac{\log(W_{2n})}{2n}\stackrel{a.s}{\longrightarrow}r^*, \text{ as } n\to\infty,$$ where 
    $r^*=\frac{1}{2}\mathbb{E}\left(\log\left(S_4\right)\right)\geq 0$, which is strictly positive whenever $a\neq0$ and $\mathbb{E}\left(\frac{1}{S_4}\right)\leq 1$.
\end{theorem}
[See Section A.4 of Supplementary for the proof]

\begin{remark}
An ergodic process is a random process where the time averages of the process tend to the appropriate ensemble averages. Formal definitions can be found in \cite{Billingsley1966ErgodicTA}. Ergodicity serves as a common and crucial assumption in time series analysis. For example, all autoregressive and moving average processes are ergodic.
    It can be shown that under AR(1) model, $\mathbb{E}\left(\frac{1}{S_4}\right)\leq 1$ holds true~\footnote{Indeed, note that $S_4$ is just a simple likelihood ratio, which always satisfies that its expectation is 1 under the null and its inverse has expectation 1 under the alternative.}. Hence, Theorem \ref{thm:cont-gen} can be regarded as a strict generalization of Theorem \ref{thm:ar-rate}.
\end{remark}

\section{EXPERIMENTAL RESULTS}
\label{sec:exp}

We present results on both simulated and real data.

\subsection{Simulation study for binary case}
In this subsection, we investigate the performance of our
test martingales for the binary case and compare it with the universal inference \cite{ramdas2022testing} and conformal inference (simple jumper algorithm) \cite{vovk2021testing,vovk2022conformal}. These two approaches have been described in Appendix \ref{existinng-meth}.

\begin{figure*}[!htb]
\centering
\begin{subfigure}{.45\textwidth}
  \centering
  \includegraphics[width=0.99\linewidth]{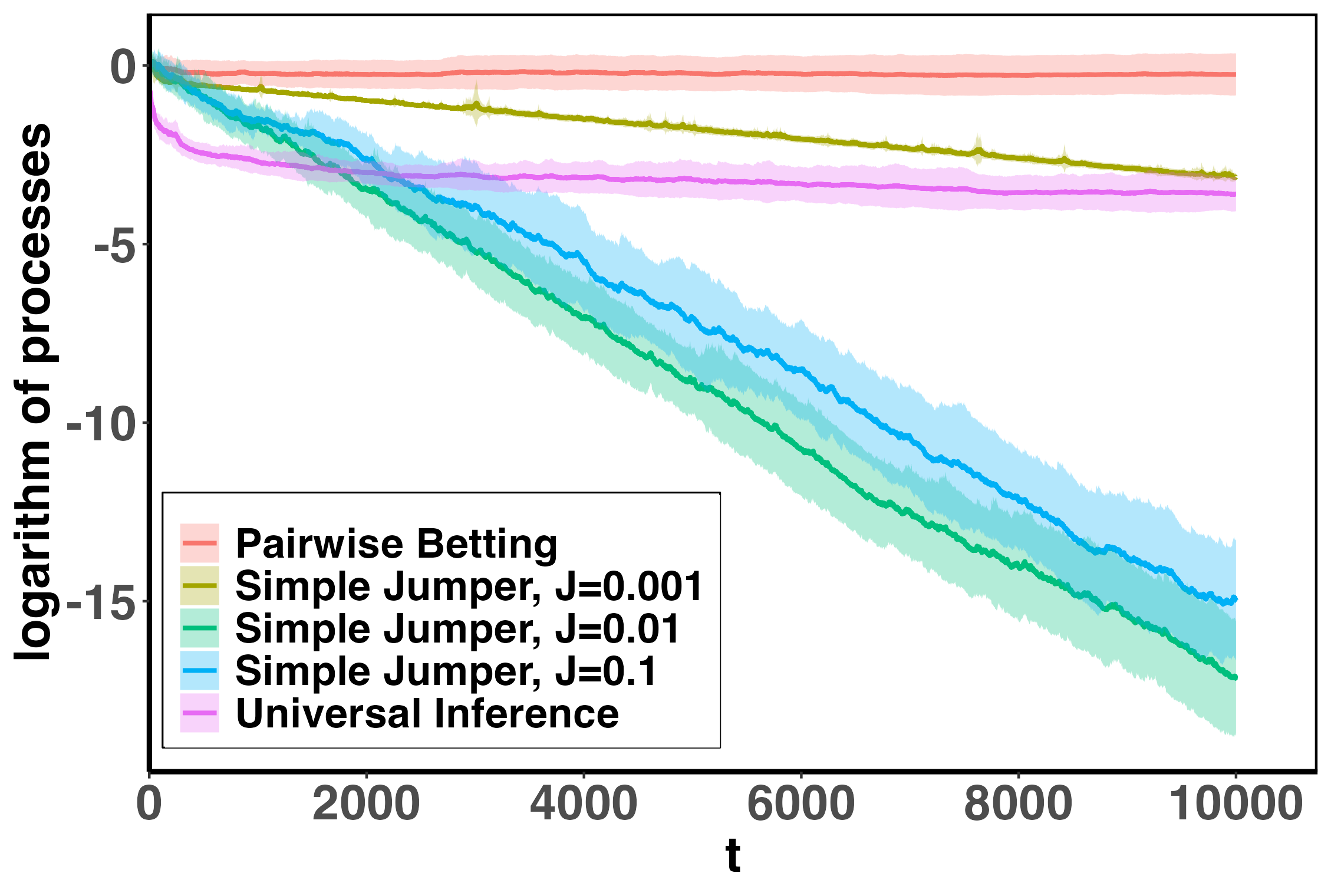}  
  \caption{Bernoulli(0.2)}
\end{subfigure}%
\begin{subfigure}{.45\textwidth}
  \centering
  \includegraphics[width=0.99\linewidth]{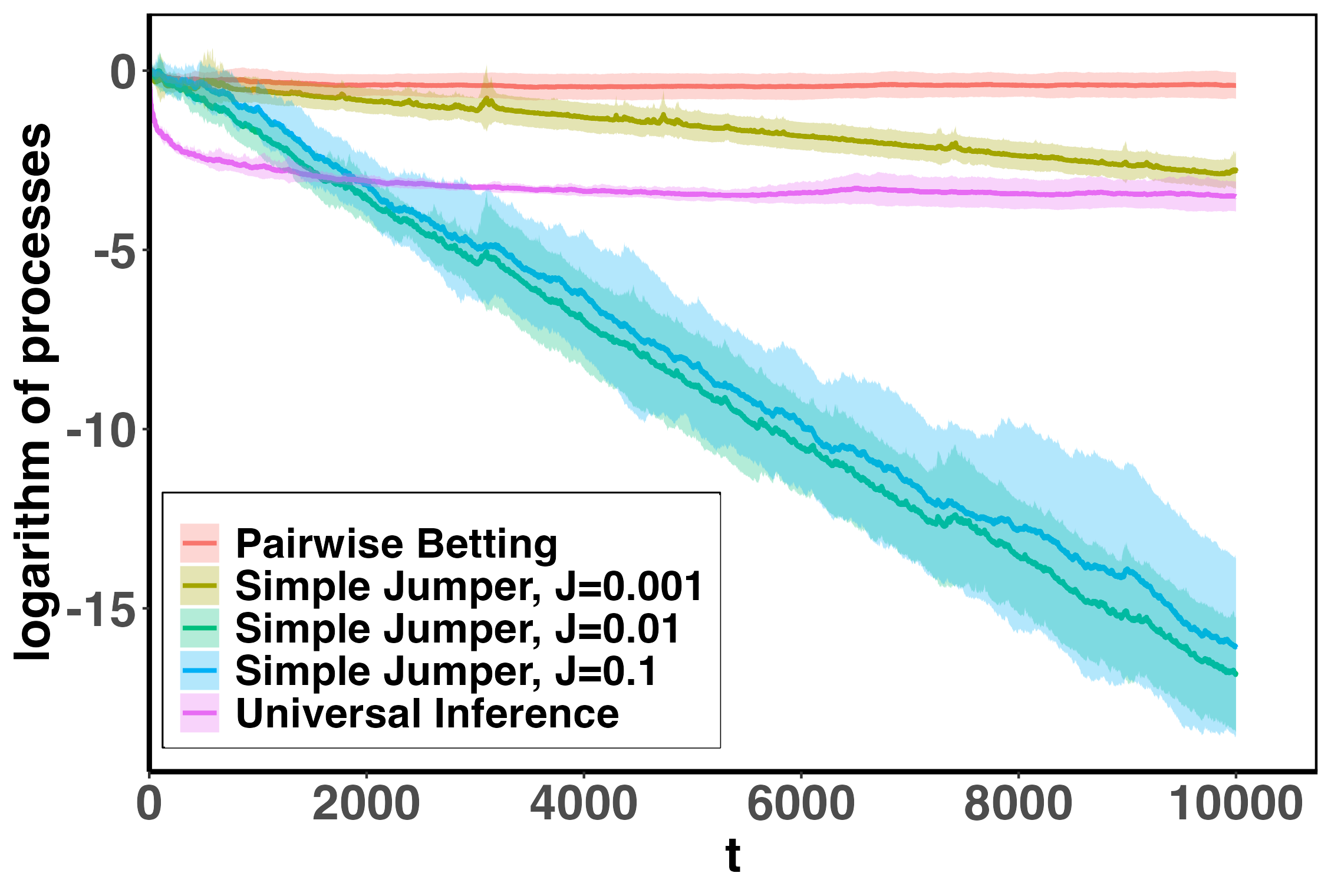}  
  \caption{Bernoulli(0.5)}
\end{subfigure}
\begin{subfigure}{.45\textwidth}
  \centering
  \includegraphics[width=0.99\linewidth]{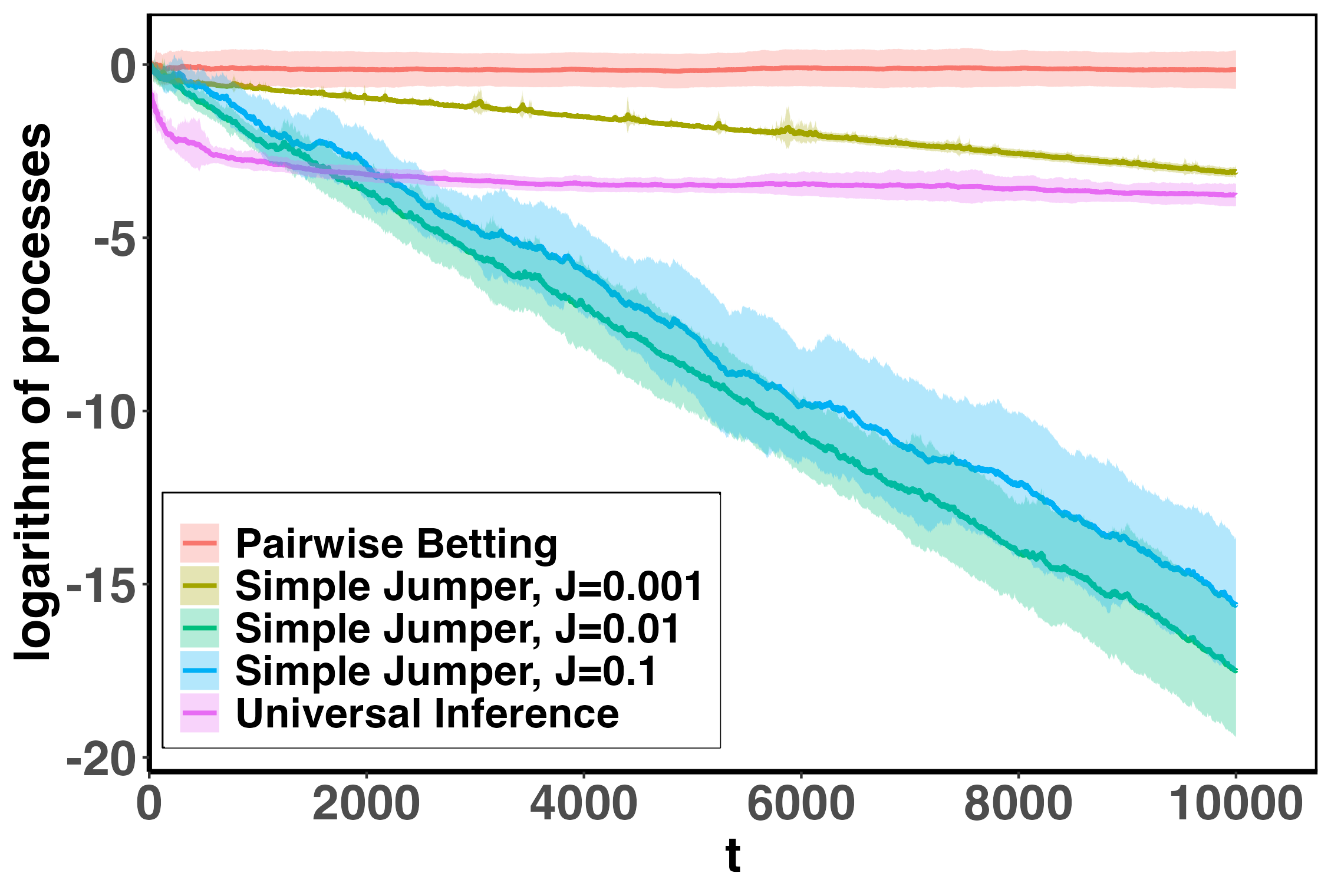}
  \caption{Bernoulli(0.8)}
\end{subfigure}
\caption{Evolution of our pairwise betting ($\log M_t$), universal inference (\cite{ramdas2022testing}) and conformal inference (simple jumper algorithm, with rate = 0.1, 0.01, 0.001) (\cite{vovk2021testing}) under iid\ Bernoulli models. The average +/- standard deviation of $10$ independent simulations is plotted. This experiment is a sanity check: as expected, none of the evidence processes grow up with time since the null is true.}
\label{fig:bern}
\end{figure*}

\paragraph{No power against iid Bernoulli.}
We conduct a sanity check to verify that our evidence measure does grow against iid Bernoulli sources (which are exchangeable).
 Our experiments encompass three specific cases: $\text{Bernoulli}(0.2)$, $\text{Bernoulli}(0.5)$ and $\text{Bernoulli}(0.8)$.
In all instances, $\log(M_t)$ does not grow with $t$; see Figure~\ref{fig:bern}.

\begin{figure*}[!htb]
\centering
\begin{subfigure}{.45\textwidth}
  \centering
  \includegraphics[width=1\linewidth]{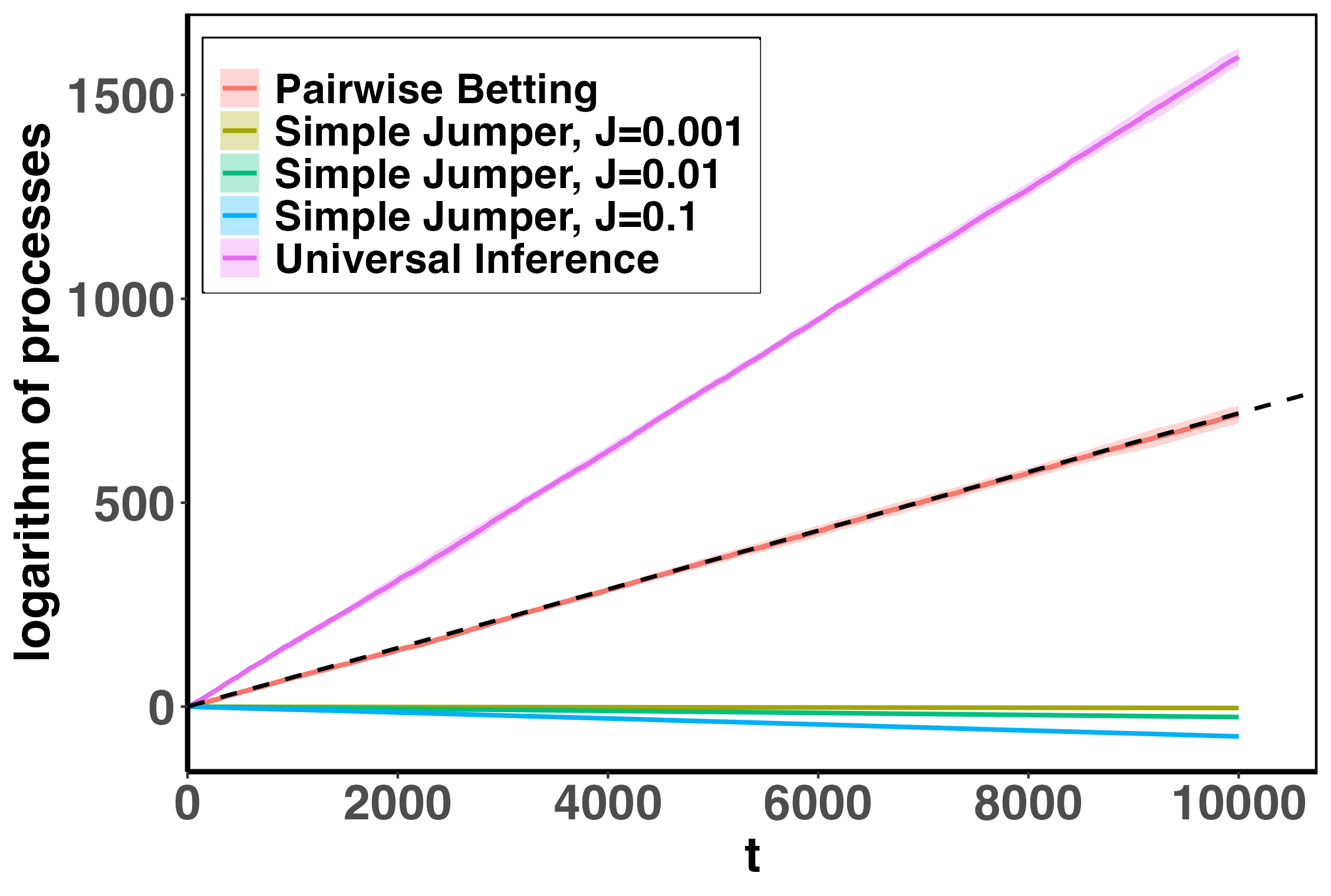}
  \caption{Markov(0.9,0.1)}
\end{subfigure}%
\begin{subfigure}{.45\textwidth}
  \centering
  \includegraphics[width=1\linewidth]{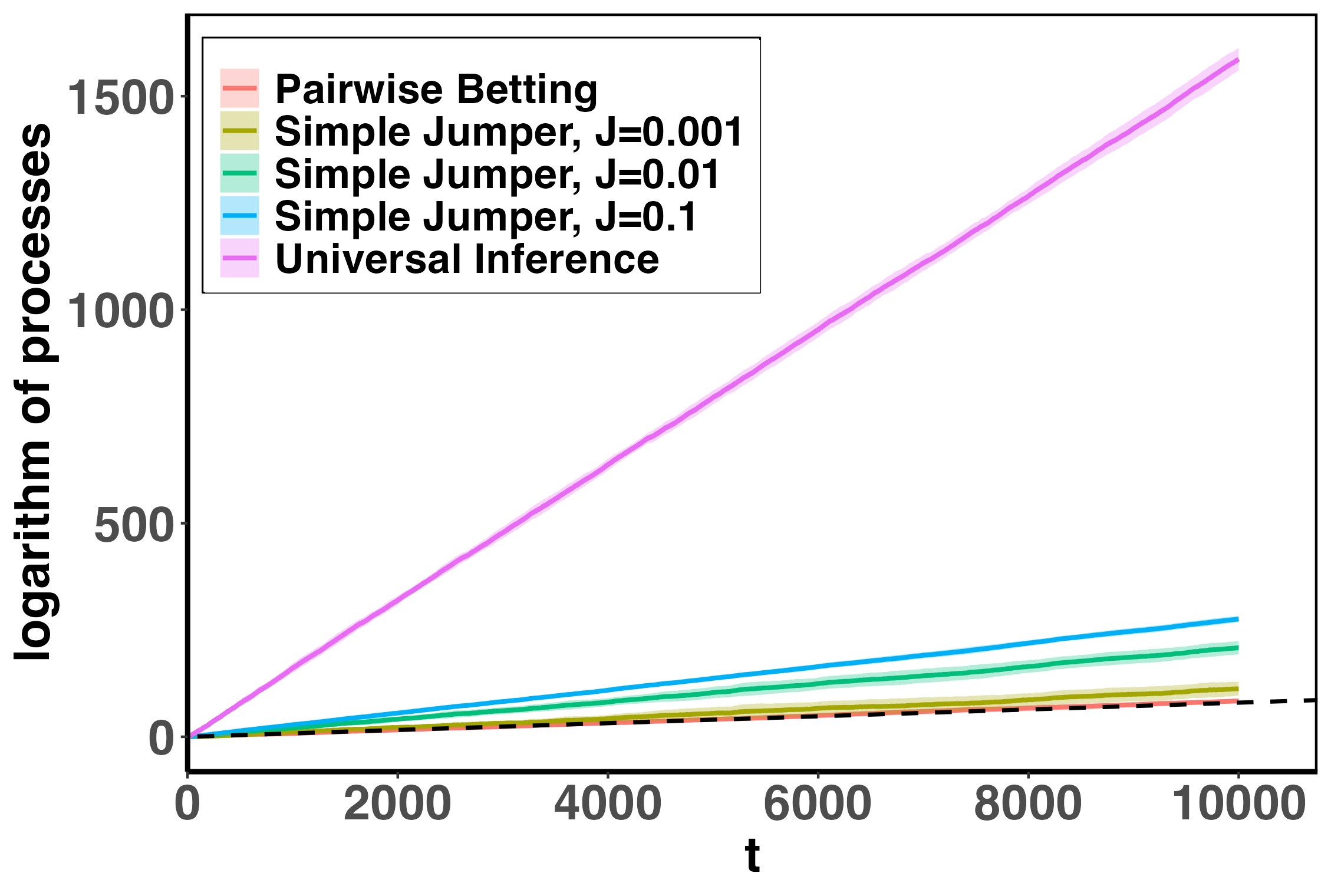}
  \caption{Markov(0.1,0.9)}
\end{subfigure}
\begin{subfigure}{.45\textwidth}
  \centering
  \includegraphics[width=1\linewidth]{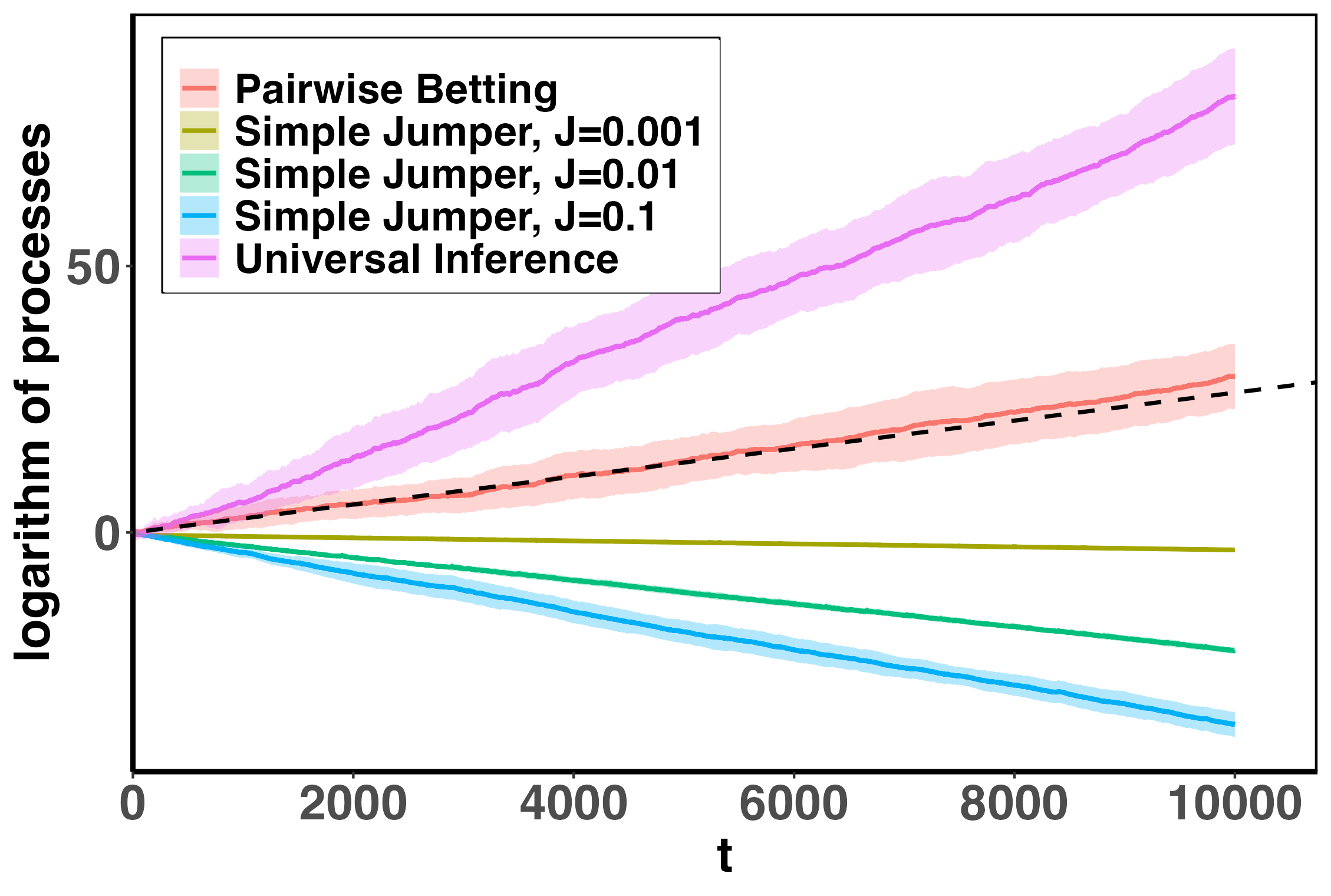}
  \caption{Markov(0.6,0.4)}
\end{subfigure}%
\begin{subfigure}{.45\textwidth}
  \centering
  \includegraphics[width=1\linewidth]{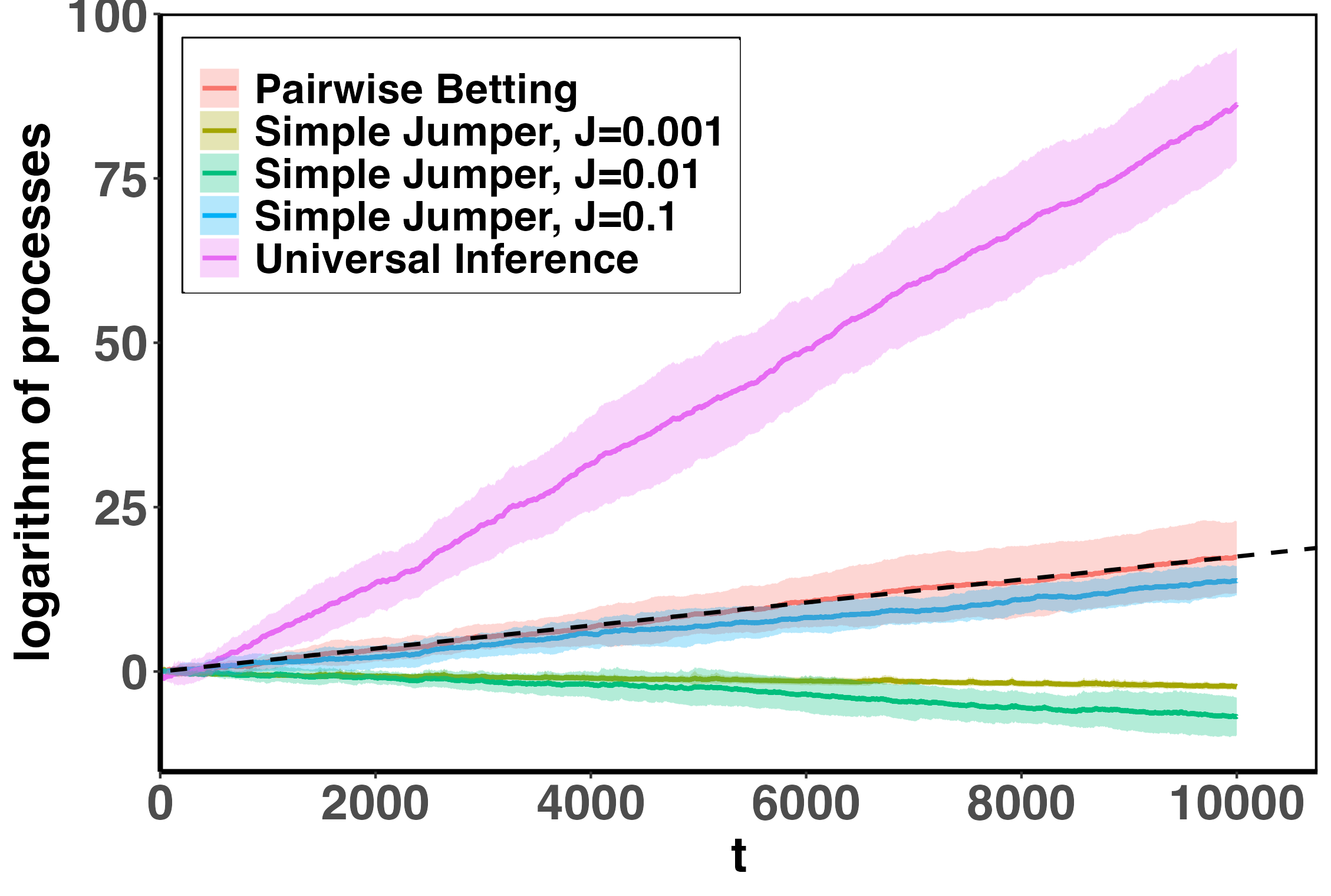}
  \caption{Markov(0.4,0.6)}
\end{subfigure}
\caption{Evolution of our pairwise betting process  ($\log M_t$), universal inference \cite{ramdas2022testing} and conformal inference (simple jumper algorithm, with rate $J=0.1,0.01,0.001$) \cite{vovk2021testing} under four different Markov Models. The average +/- standard deviation of $10$ independent simulations is plotted. The black dotted lines are the lines with slope $r$ (Theorem 2.1), which perfectly predicts the evolution of our process.}
\label{fig:Markov}
\end{figure*}

\paragraph{Power against a Markov alternative.}
The probability that the first observation is 1 is
assumed $0.5$.
In our computational experiments, we explore four specific cases: $\text{Markov}(0.4, 0.6)$, $\text{Markov}(0.6, 0.4)$, $\text{Markov}(0.1, 0.9)$ and $\text{Markov}(0.9, 0.1)$ (We use the notation $\text{Markov}(\pi_{1|0},\pi_{1|1})$ for the probability
distribution of a Markov chain with the transition probabilities $\pi_{1|0}$ and
$\pi_{1|1}$). Figure \ref{fig:Markov} shows that our theoretical black dotted line perfectly predicts practical performance, and also that in 3 out of the 4 plots, our method outperforms the conformal inference approach (for which there are no guarantees of consistency), and that although the universal inference approach is better than ours, it does not apply to continuous data settings.


\subsection{Simulation study for continuous case}
We now investigate the performance of our
test martingales for the continuous case. We have drawn the first observation from a standard normal distribution. Our computational experiments encompass five specific values of the unknown parameter $a$ of AR(1) model with a known variance of the white noise, $\sigma^2=1$. 

\begin{figure}[!htb]
    \centering
    \includegraphics[width=0.5\linewidth]{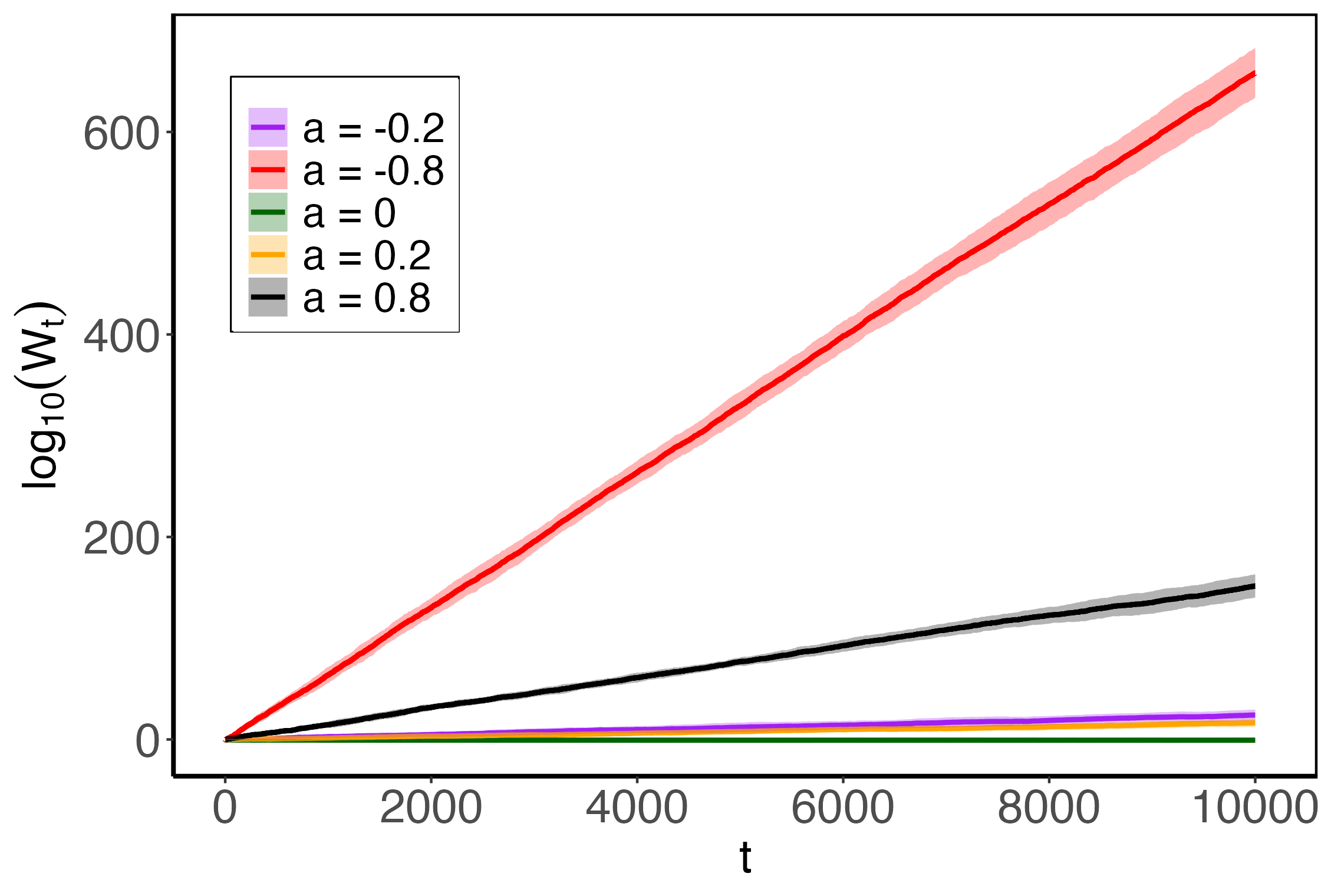}
    \caption{Average values of $\log W_t$ (+/- standard deviation) of $10$ independent simulations for AR(1) model with five different choices of the parameter $a$. Except for $a=0$ (when the null is true), $\log W_t$ grows linearly with time $t$, as predicted by our theory. Note that positive and negative $a$ behave differently, as they should. When $a>0$, all terms are positively correlated, and when $a<0$, the correlations between odd-spaced terms is negative and even-spaced terms is positive.}
    \label{fig:cont}
\end{figure}

As illustrated in Figure \ref{fig:cont}, we observe that for $a=0$ (representing the iid normal case), the logarithm of our process does not grow with time, whereas for $a=\pm 0.2, \pm 0.8$, the logarithm of our process grows linearly with time.

\subsection{Real data experiment}
\begin{figure}[!htb]
     \centering
     \includegraphics[width=0.5\linewidth]{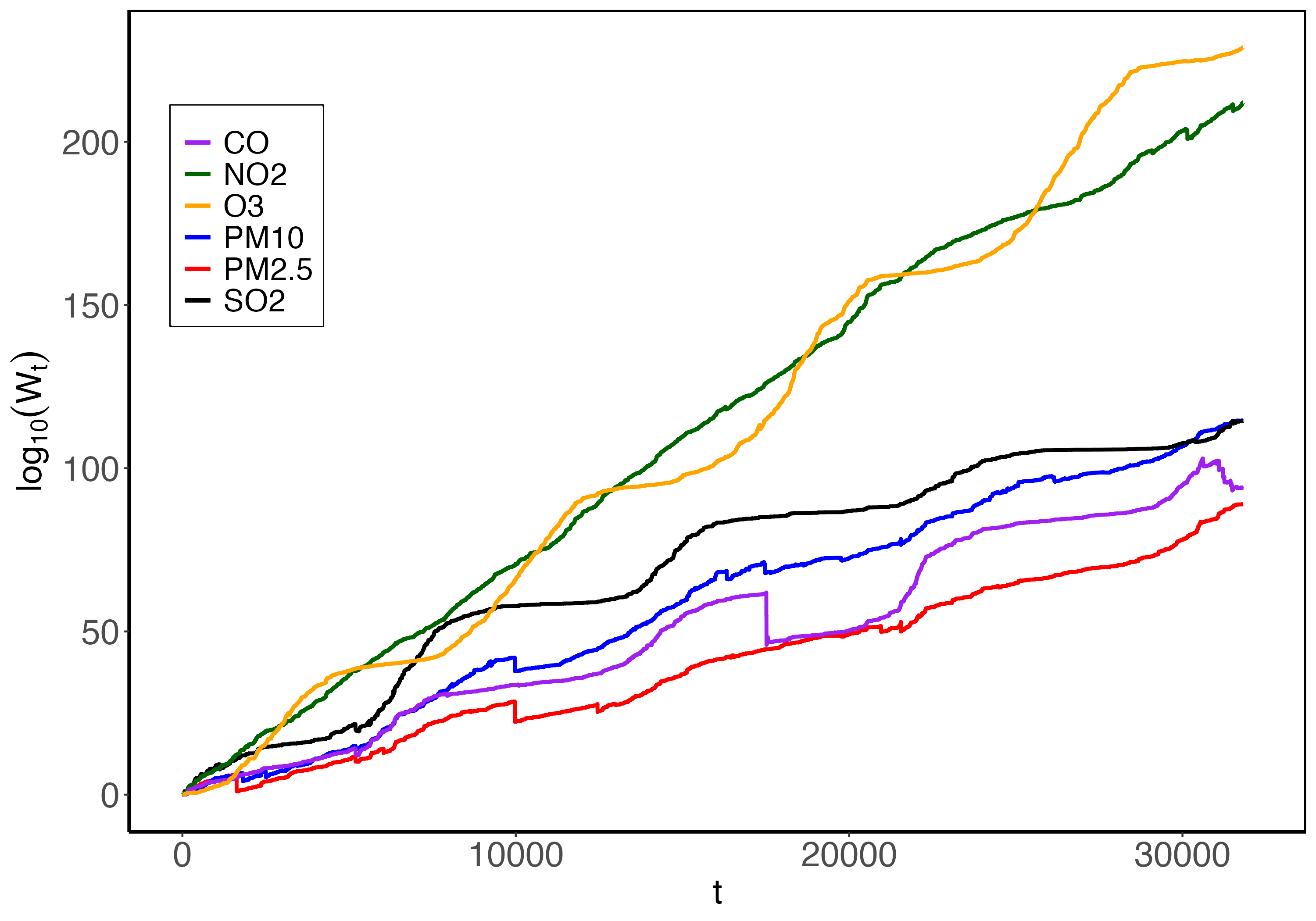}
     \caption{Accumulating evidence against the individual hypotheses that the sequences of hourly concentration of six air-pollutants are exchangeable.}
     \label{fig:airquality}
 \end{figure}

 We conclude the empirical evaluation with an implementation of our method on the Beijing Multi-Site Air-Quality Data \cite{misc_beijing_multi-site_air-quality_data_501}, which contains hourly observations of six main air pollutants over the time period from March 1, 2013 to February 28, 2017 at multiple sites in Beijing. For our analysis, we focus on the time series data from the Aotizhongxin station. Figure \ref{fig:airquality}, shows the growth of $\log(W_t)$ with $t$, which clearly indicates that the null hypothesis can be safely rejected, for all these six sequences. This empirical validation reaffirms the practical utility of our testing methodology in real-world scenarios.

 \section{DISCUSSION}
\label{sec:discussion}

 \subsection{Versatility of Our Method}

It is crucial to emphasize that in both binary and continuous cases, the models we employed under the alternative only serve as a strategic aid for betting. The model's use does not impose any constraints on our $H_0$: it just guides our betting in an attempt to prove the null false, but if the null is true, the evidence cannot grow no matter what model we use. 

The choice of a model is most impactful in terms of the power of the test. If the model does not perfectly align with the underlying data generation process, it may potentially reduce the power. However, it is essential to note that our test is ``safe" in the sense that it never compromises the control of the type-1 error.

The two specific cases (first-order Markov and AR(1)) we considered in our paper serve as illustrative examples for the sake of clarity and simplicity, offering valuable insights into the testing process due to analytical tractability of optimal bets and wealth growth rates. However, our method can be \emph{readily used in any other setting}. The conditional likelihood under null is always $\frac{1}{2}$. So, the key requirement is the availability of a conditional generative model that can be learned/updated online, to facilitate the betting process.

 \subsection{Even and odd games}

It is important to note that instead of betting in the odd time steps, as described earlier, we could bet in the even time steps as well. And it may initially seem that the average wealth of these two games might be a valid measure of evidence against null, but this turns out to not be the case, as we explain below.

Clearly, the same player cannot play both the games: betting relies on uncertainty, and it's meaningless to consider a game where the bettor wagers on something already known to them. Once a bet has been placed on $Z_{2t+1}=(X_{2t+1},X_{2t+2})$, and one has learned the outcome, it becomes meaningless to bet on $Z_{2t+2}=(X_{2t+2},X_{2t+3})$ because $X_{2t+2}$ is already known. One may think to not reveal the outcome of the first bet until after the second bet is made, but this does not work because knowing $X_{2t+1}$ is key to betting on $Z_{2t+2}$.

Additionally, the games corresponding to even and odd times yield two different filtrations, which we denote as $\mathcal{F}_1$ and $\mathcal{F}_2$, respectively. Both $\mathcal{F}_1$ and $\mathcal{F}_2$ represent coarsenings of the data filtration ($\mathcal{F}$), but in two different ways. If we denote the wealth processes in these two games as $\{M_t\}_{t\geq0}$ and $\{N_t\}_{t\geq0}$, respectively, they are test martingales with respect to $\mathcal{F}_1$ and $\mathcal{F}_2$, respectively. Consequently, the object $T = \frac{M+N}{2}$ is not a test martingale because it is neither adapted to $\mathcal F_1$ nor to $\mathcal F_2$, and it doesn't seem to be an e-process either. While it is adapted to $\mathcal{F}$, it does not qualify as a test martingale with respect to $\mathcal{F}$ (indeed, these do not exist.)

It is indeed intriguing to consider the average wealth obtained by \emph{two independent, non-communicating players} engaging in two different games with nature. It may be interesting for future work to investigate its properties.

\subsection{Pairwise Betting: A Broader Outlook}

The absence of a powerful test martingale in the original data filtration is a phenomenon that is encountered in other significant nonparametric hypothesis classes. For example, for the fundamental problem of independence testing, \cite{henzi2023rank} shows that test martingales are powerless. However, \cite{podkopaev2023sequential} showed that a (different) pairwise betting strategy yields a consistent and powerful test martingale for the problem, while \cite{henzi2023rank} reduced the filtration in a different fashion analogous to conformal prediction.

Thus, there are interesting parallels between the situations encountered for testing independence and for testing exchangeability. These two problems are definitely related; in nonsequential settings, most tests for independence proceed via testing exchangeability using a permutation test. But the relationship in the sequential setting is complicated by the fact that testing independence can occur in non-iid settings as well, as was demonstrated in~\cite{podkopaev2023sequential}. 


\subsection{Testing with more than two observations together}
Our strategy involves betting on pairs of observations due to the powerlessness of betting on individual observations, but one also has the flexibility to process more than two observations at a time.
The obvious cost of doing this is further shrinkage of filtration, meaning that optional stopping is effectively constrained to only stop at times divisible by the number of data points processed simultaneously. However, utilizing larger batches of data points might improve the growth rate of the wealth. For example, a detailed analysis of testing by betting with three consecutive observations is in Appendix~\ref{betting-triple}.
 
\section{CONCLUSION}
\label{conc}
Our paper introduces a novel approach to the fundamental question of sequentially testing exchangeability, centered around the new (yet simple) idea of pairwise betting, which leads to a
nontrivial test martingale. Importantly, our method applies to any general observation space and is amenable to analytical study.
We have provided a detailed analysis of our approach for both binary and continuous cases, specifically focusing on Markov and AR(1) alternatives (but extended to a broader class of alternatives), respectively, for which we demonstrated the consistency of our approach and explored its growth rate.

\subsubsection*{Acknowledgments}
AR acknowledges support from NSF IIS-2229881 and NSF DMS-2310718.

\newpage
\appendix


\section{Details of existing methods}
\label{existinng-meth}

\textbf{Universal Inference based approach:} \cite{ramdas2022testing} has used universal inference \cite{wasserman2020universal}, incorporating the method of mixtures with Jeffreys’ prior, to handle the composite alternative, along with the maximum likelihood under the null, to ultimately yield a computationally efficient closed-form e-process
$$
R_n:=\frac{\Gamma\left(n_{0 \mid 0}+\frac{1}{2}\right) \Gamma\left(n_{0 \mid 1}+\frac{1}{2}\right) \Gamma\left(n_{1 \mid 0}+\frac{1}{2}\right) \Gamma\left(n_{1 \mid 1}+\frac{1}{2}\right)}{2 \Gamma\left(\frac{1}{2}\right)^4 \Gamma\left(n_{0 \mid 0}+n_{1 \mid 0}+1\right) \Gamma\left(n_{0 \mid 1}+n_{1 \mid 1}+1\right)} /\left(\left(\frac{n_1}{n}\right)^{n_1}\left(\frac{n_0}{n}\right)^{n_0}\right);~ n=1,2,3,\cdots.
$$
Here $\Gamma$ denotes the usual gamma function, $n_i$ is the number of times $i$ has observed among first $n$ observations and $n_{i|j}$ is the number of transitions from $j$ to $i$ observed upto $n$-th observations, for $i,j=1,2$.
This process is not a test supermartingale, but is upper-bounded by some nonnegative martingale for every exchangeable distribution, and thresholding it at level $1/\alpha$ yields a level $\alpha$ sequential test for exchangeability. 

\textbf{Conformal Inference based approach:}
\cite{vovk2021testing} introduced a method for testing exchangeability based on conformal prediction, wherein the canonical data filtration is replaced by a less informative filtration composed of conformal p-values. In this approach, a sequence of independent conformal p-variables are generated under the null, which are transformed into a test martingale through suitable calibration. The idea is to bet
against the uniform distribution of the conformal p-values. The Simple Jumper Algorithm \cite{vovk2021retrain} is one such method, which takes the conformal p-values as input and produces a conformal test martingale as output. It also involves a hyperparameter, $J \in (0,1)$. The method is briefly described below.

The conformal p-values $p_1, p_2, \ldots$ are transformed into conformal test martingale as follows:

\begin{equation}
\label{calib}
F\left(p_1, \ldots, p_n\right):=\int\left(\prod_{i=1}^n f_{\epsilon_i}\left(p_i\right)\right) \mu\left(\mathrm{d}\left(\epsilon_0, \epsilon_1, \ldots\right)\right),
\end{equation}
where
\begin{equation}
\label{lin-calib}
  f_\epsilon(p):=1+\epsilon(p-0.5) , 
\end{equation}
and $\mu$ denotes the following Markov chain with state space $\{-1,0,1\}$: the initial state is $\epsilon_0\in\{-1,0,1\}$ with equal probabilities, and the transition function prescribes maintaining the same state with probability $1-J$ and, with probability $J$, choosing a random state from the state space $\{-1,0,1\}$. The intuition is that at each step $i$, one of the betting functions \ref{lin-calib} is used: $f_{-1}$ corresponds to betting on small values of $p_i, f_1$ corresponds to betting on large values of $p_i$, and $f_0$ corresponds to not betting.

\section{Testing by betting with three consecutive observations}
\label{betting-triple}
Instead of processing the data sequence pairwise, we now consider three consecutive observations together. 
This new betting game leads to a test martingale, which appears to have a higher growth rate than the existing one in many situations. Notably, the introduced modification comes with a minor trade-off: optional stopping is now permitted only at times divisible by three, in contrast to the even times permitted in pairwise betting.

\subsection{Test for Binary Observations}
Consider a sequence of binary random variables
$X_1,X_2,\cdots$. The realization of the random variables
are denoted as $x_1,x_2,\cdots$. We primarily focus on first-order
Markov alternative, as done before.

For a betting game starting with an initial wealth of $M^*_{0}=1$, we define $V_t=\{X_{3t+1}, X_{3t+2}, X_{3t+3}\}$ for $t=0,1,,\cdots$ to be the \emph{unordered} set of three consecutive random variables. If $V_t=\{0,0,0\}$ or $\{1,1,1\}$, there is nothing to bet for, and we denote this event as $A_t$. Otherwise, we place bets on the order in which they occur, according to the likelihood ratio, conditioned on the observed values of $\mathbf{X}^{3t}:=\{X_1,\cdots,X_{3t}\}$ and the unordered set $V_t$ and the event $A_t^c$. Thereafter, nature unveils the observed values of $X_{3t+1}, X_{3t+2}$ and $X_{3t+3}$, which is some permutation of the elements of $V_t$. Notably
the conditional likelihood under $H_0$ is $1/3$, since all the permutations are equally likely under the null of exchangeability, and given that all three binary observations are not equal, two of them must be equal, implying there are three distinct permutations (we dente them by $\pi_{t,i},~ i=1,2,3$). And the conditional likelihood under $H_1$ is
\begin{align}
\label{alt-lik}
&\nonumber\mathbb{P}_{H_1}(X_{3t+1}=x_{3t+1}, X_{3t+2}=x_{3t+2}, X_{3t+3}=x_{3t+3}|\mathbf{X^{3t}}=
                             \mathbf{x^{3t}},V_t=v_t,A^c_t)\\=&\frac{h(x_{3t},x_{3t+1},x_{3t+2},x_{3t+3})}{\sum_{i=1}^3h(x_{3t},x^{i}_{t,1},x^{i}_{t,2},x^{i}_{t,3})},
\end{align}
where $h(w,x,y,z):={p}_{z|y}{p}_{y|x}{p}_{x|w}$ and $(x^{i}_{t,1},x^{i}_{t,2},x^{i}_{t,3})$ is the ordered set corresponding to the permutation $\pi_{t,i}$ of the unordered set $v_t$. Here, ${p }_{i|j}$ is the transition probability from $j$ to $i$ of the underlying Markov model.
Then, the betting score at $t$-th
round of betting for the Oracle (denoted as $B^*_t$) is the
likelihood ratio of the alternative (Equation \eqref{alt-lik}) to the null (which is $\frac{1}{3}$), i.e., $B^*_t=\mathbbm{1}_{A_t}+\frac{3h(X_{3t},X_{3t+1},X_{3t+2},X_{3t+3})}{\sum_{i=1}^3h(X_{3t},X^{i}_{t,1},X^{i}_{t,2},X^{i}_{t,3})}\mathbbm{1}_{A^c_t}$.

But, in practice, we replace ${p}_{i|j}$ (which is unknown) by its maximum likelihood estimator (MLE) based on the first $3t$ many observations (denoted by $\hat{p}_{i|j}$) in Equation \eqref{alt-lik}. But there could be other choices too.

We do not bet on the first round, i.e., $\hat{B^*_1}=1$. And betting
score at $t$-th round of betting is
\begin{equation}
\hat{B}^*_t=\mathbbm{1}_{A_t}+\frac{3\hat{h}_t(X_{3t},X_{3t+1},X_{3t+2},X_{3t+3})}{\sum_{i=1}^3\hat{h}_t(X_{3t},X^{i}_{t,1},X^{i}_{t,2},X^{i}_{t,3})}\mathbbm{1}_{A^c_t}~; ~ t=2,3,\cdots.
\end{equation}
Here, $\hat{h}_t$ is the plug-in MLE of $h$, that is obtained by
replacing ${p}_{i|j}$ by $\hat{p}_{i|j}$ in the expression of $\hat h$, for $i,j=1,2$ and $k=1,2,3$. Thus, the bettor’s wealth after $t$ rounds of betting is

\begin{equation}
\label{a-eq:test}
M^*_{t}=M^*_{t-1}\times \hat{B}^*_{t}=\prod_{i=1}^t \hat{B}^*_{i};~t=1,2,3,\cdots.
\end{equation}
It is easy to check that $\{M^*_t\}_{t\in\mathbb{N}}$ is a test martingale for $H_0$. Hence, 
$\tau:=\inf\{3t : M^*_{t}\geq 1/\alpha,~ t\in \mathbb N\} $
  is the stopping time at which we reject the null, yielding a level $\alpha$ sequential test. 

\begin{theorem}
\label{thm:mc-triple}
Under the first-order Markov alternative, assume that $p_{0|1},p_{1|0}\neq 0$ or $1$. Then,
$\frac{\log (M^*_{n})}{3n}\to \Tilde{r}'$ almost surely as $t\to \infty$, where
$$\Tilde{r}'=\frac{1}{3}\sum_{i=0}^1\sum_{\substack{j,k,l=0\\(j,k,l)\neq\\(0,0,0),(1,1,1)}}^1\log\left(\frac{3p_{j|i}p_{k|j}p_{l|k}}{\sum_{\pi \in \Pi(j,k,l)} p_{\pi(j)|i}p_{\pi(k)|\pi(j)}p_{\pi(l)|\pi(k)}}\right) p_ip_{j|i}p_{k|j}p_{l|k}.$$ 
  Here $\Pi(j,k,l)$ is the set of all distinct permutations of the numbers $j,k,l$ and $p_i$ is the stationary probability of $i$-th state. Further, $\Tilde{r}' = 0$ 
  if $p_{1|0}=p_{1|1}$, and $\Tilde{r}' > 0$ otherwise. 
\end{theorem}

This theorem shows the consistency of our sequential level $\alpha$ test, by proving that our test martingale increases to infinity
exponentially fast in $n$. The result is quite similar to the result for the pairwise betting approach.

\subsection{Test for Continuous Observations}
Consider a sequence of continuous random variables
$X_1,X_2,\cdots$. The realization of the random variables
is denoted as $x_1,x_2,\cdots$. Our focus is on first-order
Gaussian autoregressive alternative, as before :
\begin{align*}
H_1 & : X \text{ is a stationary AR(1) process, with }  X_{t+1}=aX_{t}+\varepsilon_{t+1},
~t=1,2,\cdots,~ \varepsilon_t\stackrel{i.i.d}{\sim} N(0,\sigma^2).
\end{align*}
Here $a$ and $\sigma$ are unknown.
Using the same idea that we employed in the binary case, we obtain the test martingale. The only difference is that for continuous random variables, the probability that two random variables are equal is zero. Hence, the likelihood under $H_0$ becomes $1/6$, since there are $3!$ distinct permutations (denote them by $\tau_{t,i}$, $i=1,\cdots,6$). Similarly, as before, conditional likelihood under alternative
\begin{align}
&\nonumber\mathbb{P}_{H_1}(X_{3t+1}=x_{3t+1}, X_{3t+2}=x_{3t+2}, X_{3t+3}=x_{3t+3}|\mathbf{X^{3t}}
                           =\mathbf{x^{3t}},V_t=v_t)\\&=\frac{g(x_{3t},x_{3t+1},x_{3t+2},x_{3t+3})}{\sum_{i=1}^6g(x_{3t},x^{i}_{t,1},x^{i}_{t,2},x^{i}_{t,3})},   
\end{align}
with $g(w,x,y,z):=\frac{1}{2\pi\sigma^2}\exp{\left(-\frac{1}{2\sigma^2}(x-aw)^2-\frac{1}{2\sigma^2}(y-ax)^2-\frac{1}{2\sigma^2}(z-ay)^2\right)}$ and $(x^{i}_{t,1},x^{i}_{t,2},x^{i}_{t,3})$ is the ordered set corresponding to the permutation $\tau_{t,i}$ of the unordered set $v_t$.
So, betting score for Oracle at $t(\geq 2)$-th round is $S^*_t=\frac{6g(X_{3t},X_{3t+1},X_{3t+2},X_{3t+3})}{\sum_{i=1}^6g(X_{3t},X^{i}_{t,1},X^{i}_{t,2},X^{i}_{t,3})}$. Recall that we don't bet at $t=1$, so $S^*_1=1$.
For practical use, we need to estimate the parameters $a$ and $\sigma^2$. One viable option is to 
estimate $g$ by $\hat{g}_t$, which is obtained by replacing the model parameter $a$ by its least squares estimator, $\hat{a}_{t}=\sum_{i=2}^{3t}X_iX_{i-1}/\sum_{i=1}^{3t-1}X_i^2$ and $\sigma^2$ by 
$\hat{\sigma}^2_{t}=\frac{1}{3t-1}\sum_{i=2}^{3t}(X_{i}-\hat{a}_{t-1}X_{i-1})^2$.
So, the bet at $t$-th round ($t=2,3,\cdots$) is
\begin{equation}
\hat{S}^*_{t}=\frac{6\hat{g}_t(X_{3t},X_{3t+1},X_{3t+2},X_{3t+3})}{\sum_{i=1}^6\hat{g}_t(X_{3t},X^{i}_{t,1},X^{i}_{t,2},X^{i}_{t,3})}.
\end{equation}
Thus, starting with an initial wealth of $W^*_0=1$, the bettor’s wealth after $t$ rounds of betting is \begin{equation}
\label{a-eq:test-martingale-cont}
    W^*_{t}=W^*_{t-1}\times \hat{S}^*_{t}=\prod_{i=1}^{t} \hat{S}^*_{i}.
\end{equation}
It is easy to check that $\{W^*_t\}_{t\in\mathbb{N}}$ is a test martingale, (with respect to a shrunk filtration) for testing exchangeability and
   $\tau'=\inf\{3t : W^*_{t}\geq 1/\alpha; t\in\mathbb{N}\}$ is a level $\alpha$ sequential test.

\begin{theorem}
\label{thm:ar-triple}
Let, $\{X_t\}_t$ be a stationary Gaussian AR(1) process. Then, $\frac{\log(W^*_{2n})}{2n}\stackrel{a.s}{\longrightarrow}\Tilde{r}^*, \text{ as } n\to\infty,$ where 
$\Tilde{r}^*=\frac{1}{3}\mathbb{E}\left(\log\left(S^*_2\right)\right)\geq 0$ and equality holds if and only if $a=0$ (in which case the null would be true).
\end{theorem}

This theorem shows the consistency of our sequential level $\alpha$ test, by proving that our test martingale
increases to infinity exponentially fast in n. Although the result is quite similar to the result for the
pairwise betting approach, the growth rate r of this new test martingale is
higher, as demonstrated by our simulation studies in the next section.

\subsection{Experimental Results}

In this subsection, we investigate the performance of our
test martingales for the binary case and compare it with the universal inference \cite{ramdas2022testing} and pairwise betting.

\begin{figure*}[h!]
\centering
\centering
\subfloat[$\text{Markov}(0.9,0.1)$]{\includegraphics[width=0.45\linewidth,height=0.3\linewidth]{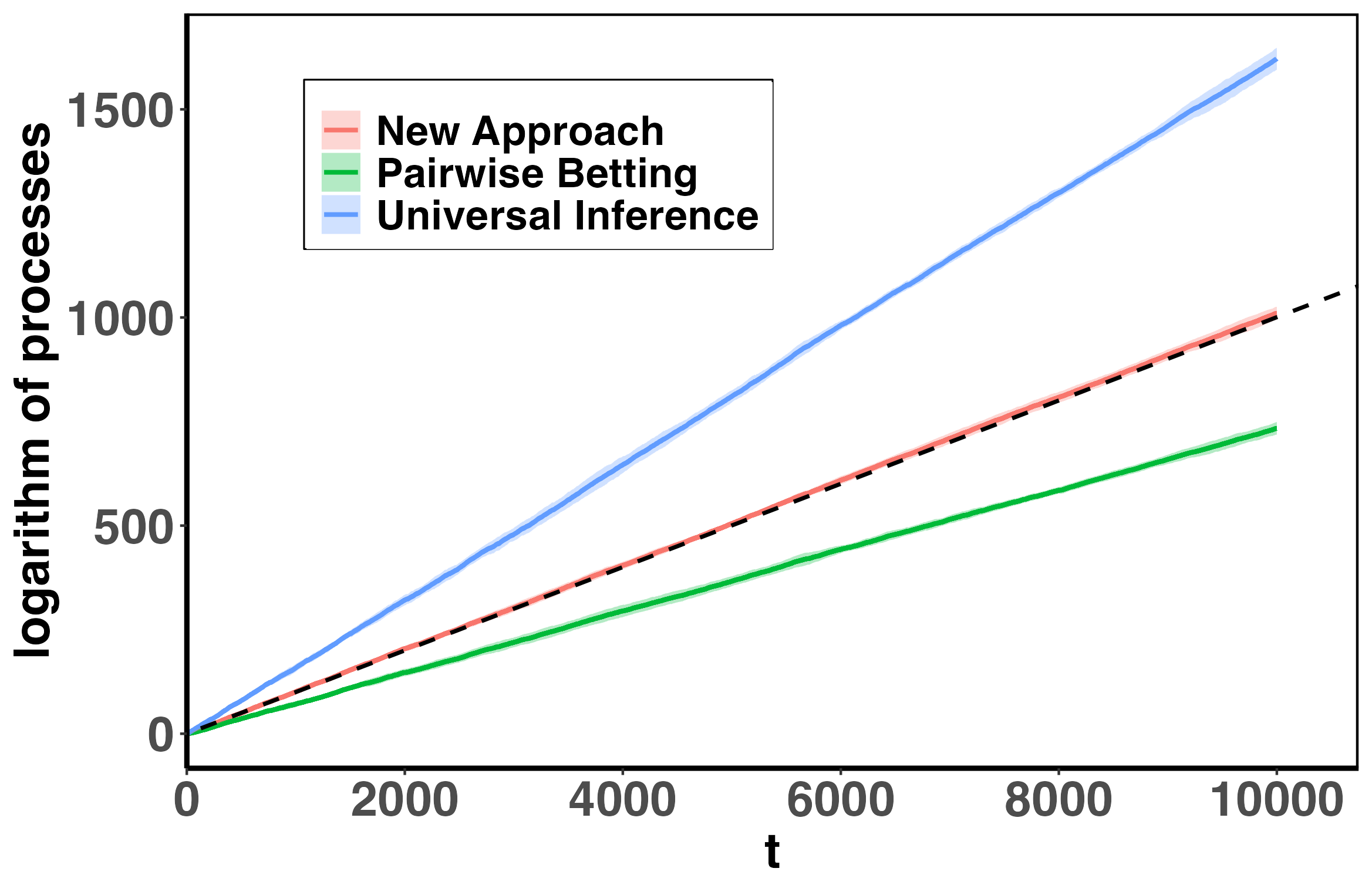}} 
\subfloat[$\text{Markov}(0.6,0.4)$]{\includegraphics[width=0.45\linewidth,height=0.3\linewidth]{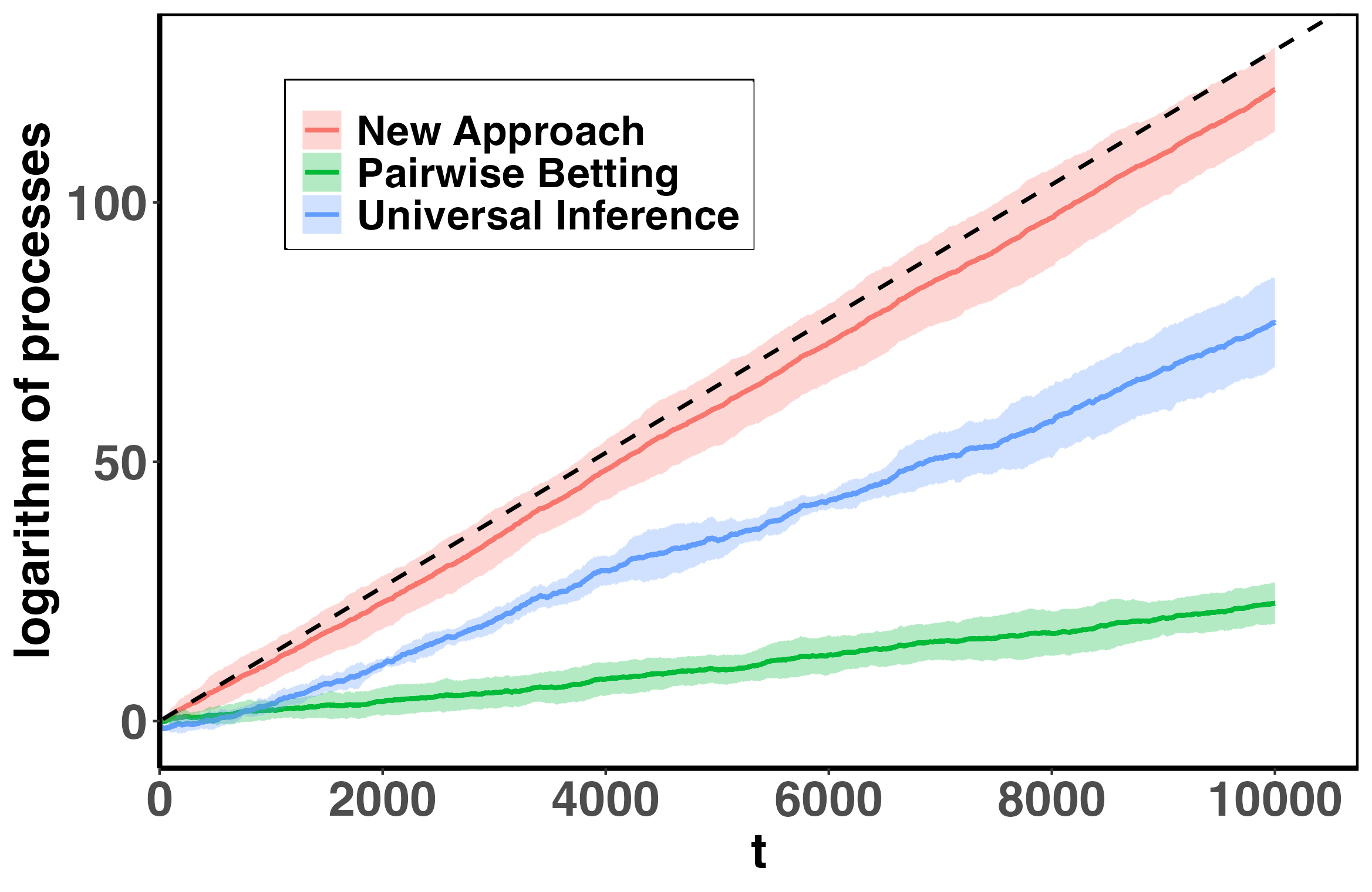}}
\caption[]{Evolution of our new wealth process based on betting with three observations, pairwise betting process  and the Universal Inference approach \cite{ramdas2022testing} for the Markov model (with four different choices of the parameters). The black dotted lines have slope $\Tilde{r}'$ (Theorem 2.1), which perfectly predicts the evolution of our process.}
\label{a-fig:Markov}  
\end{figure*}

\textbf{Simulation study for binary case:}
Figure \ref{a-fig:Markov} shows that our theoretical black dotted line perfectly predicts practical performance, and also that in both cases, our approach is better than the pairwise betting and in one case, it is better than universal inference approach.

\begin{figure*}[!htb]

\centering
\centering
\subfloat[$a=-0.8$]{\includegraphics[width=0.45\linewidth,height=0.3\linewidth]{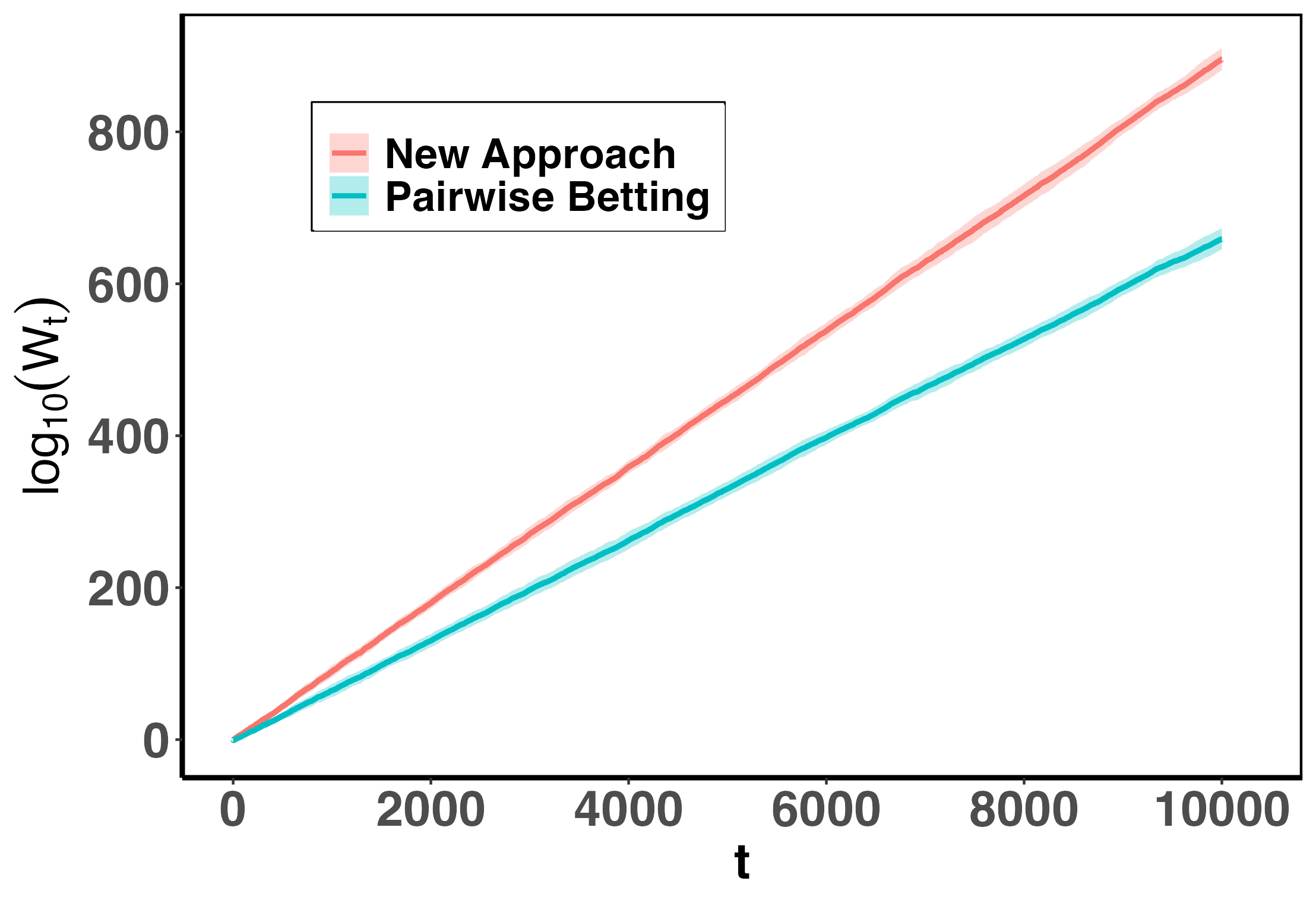}} 
\subfloat[$a=-0.2$]{\includegraphics[width=0.45\linewidth,height=0.3\linewidth]{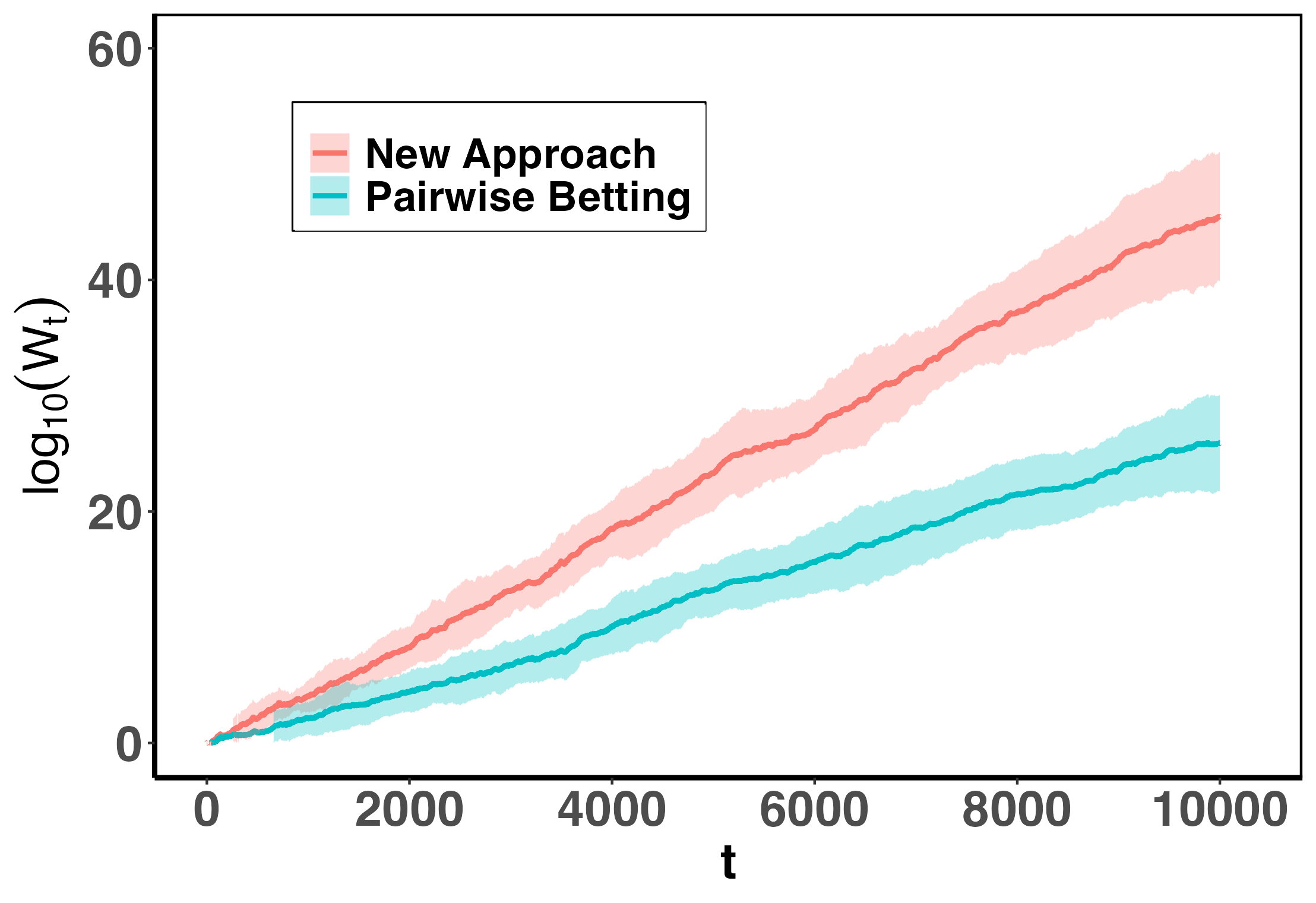}}

\subfloat[$a=0.2$]{\includegraphics[width=0.45\linewidth,height=0.3\linewidth]{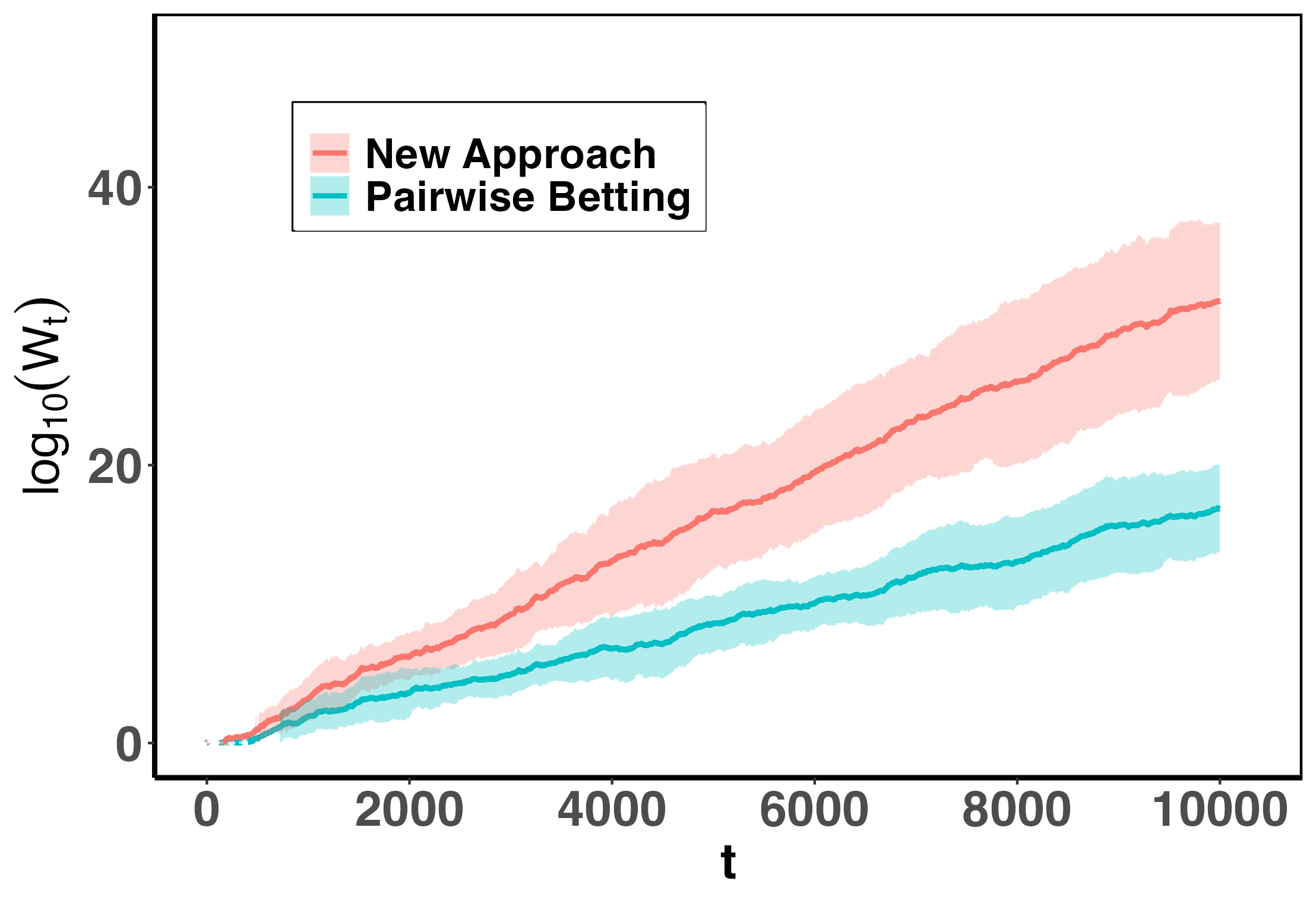}}
\subfloat[$a=0.8$]{\includegraphics[width=0.45\linewidth,height=0.3\linewidth]{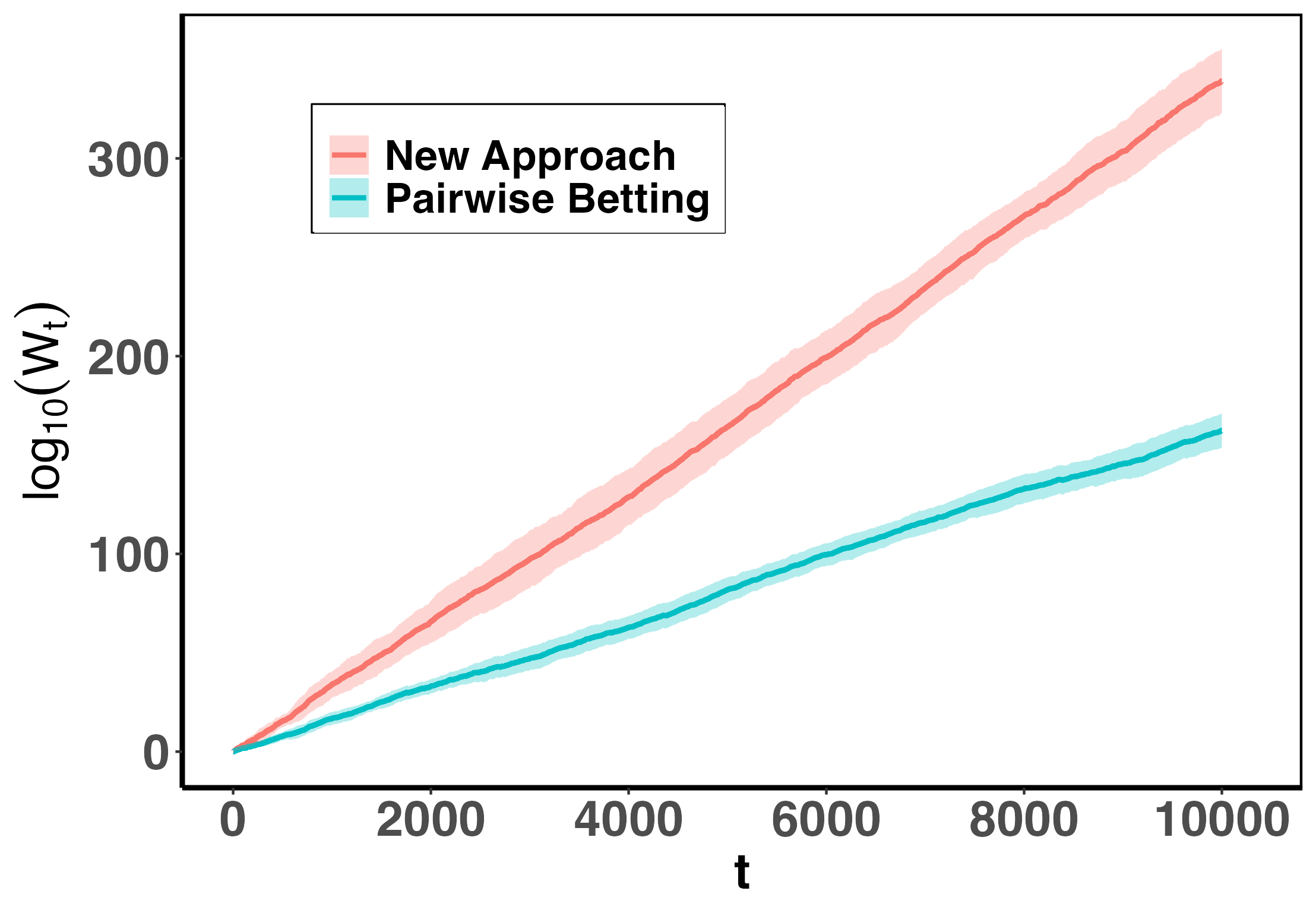}}\\
\caption[]{Evolution of the log-wealth of our new approach (betting on three observations) and the pairwise betting process for the AR(1) model (with four choices of parameter $a$). The log-wealths grow linearly as predicted by theory, but betting on triples is more powerful.}
\label{a-fig:cont}  
\end{figure*}

\textbf{Simulation study for continuous case:}
We now investigate the performance of our
test martingales for the continuous case. We have drawn the first observation from the standard normal distribution. Our computational experiments encompass five specific values of the unknown parameter $a$ of AR(1) model with a known variance of the white noise, $\sigma^2=1$. As illustrated in Figure \ref{a-fig:cont}, we observe that for $a=\pm 0.2, \pm 0.8$, the logarithm of our process grows linearly with time. In all these examples, our process demonstrates a higher growth rate than the pairwise betting approach.

\section{OMITTED PROOFS}
\label{proofs}

Here, we present the formal proofs of the theorems stated and discussed in the main paper. The key mathematical tool that we employ to establish these proofs is the ergodic theorem \cite{Billingsley1966ErgodicTA}.

\subsection{Proof of Theorem \ref{thm:rate}}

For $t\geq 2$, let us define $Y_{2t}=\{X_{2t-2},X_{2t-1},X_{2t}\}$, $\hat{D}_{2t}=\log \hat{B}_{2t}$. So, we can write $\log(M_{2n})=\sum_{t=2}^n\hat{D}_{2t}$.

Now, by using ergodic theorem, we can show that the MLE of the transition probability from $i$ th state to $j$ th state, $\hat{p}_{j|i}$ is strongly consistent for ${p}_{j|i}$, i.e,
\begin{equation}
\label{trans-p-conv}
    \hat{p}_{j|i}(t)\to {p}_{j|i}, \text{ almost surely, as } t \to \infty, \text{ for } i,j\in\{0,1\}.
\end{equation}
And since $X_1,X_2,\cdots$ is a Markov chain of first order, $Y_2,Y_4,\cdots$ is also a Markov chain of first order. So, again by ergodic theorem,
\begin{equation}
\label{ind-conv}
    \frac{1}{n-1}\sum_{t=2}^{n}\mathbbm{I}_{Y_{2t}=\{i,j,{j}^c\}}\to p_ip_{j|i}p_{j^c|j}, \text{ almost surely, as } n \to \infty, \text{ for } i,j\in\{0,1\},
\end{equation}
where $p_i=\frac{p_{i|i^c}}{p_{i|i^c}+p_{i^c|i}}.$ is the stationary probability of $i$th state. Now,
\begin{align*}
    &\frac{1}{n-1}\sum_{t=2}^{n}\hat{D}_{2t}\\
    &=\frac{1}{n-1}\sum_{t=2}^{n} \sum_{i=0}^1\sum_{j=0}^1 \log\left(\frac{2\hat{p}_{j|i}(2t)\hat{p}_{j^c|j}(2t)}{\hat{p}_{j|i}(2t)\hat{p}_{j^c|j}(2t)+\hat{p}_{j^c|i}(2t)\hat{p}_{j|j^c}(2t)}\right)\mathbbm{I}_{Y_{2t}=\{i,j,{j}^c\}}\\
    &= \sum_{i=0}^1\sum_{j=0}^1\frac{1}{n-1}\sum_{t=2}^{n} \left[\log\left(\frac{2\hat{p}_{j|i}(2t)\hat{p}_{j^c|j}(2t)}{\hat{p}_{j|i}(2t)\hat{p}_{j^c|j}(2t)+\hat{p}_{j^c|i}(2t)\hat{p}_{j|j^c}(2t)}\right)-\log\left(\frac{2{p}_{j|i}{p}_{j^c|j}}{{p}_{j|i}{p}_{j^c|j}+{p}_{j^c|i}{p}_{j|j^c}}\right)\right]\mathbbm{I}_{Y_{2t}=\{i,j,{j}^c\}}\\
    &~~+\sum_{i=0}^1\sum_{j=0}^1\log\left(\frac{2{p}_{j|i}{p}_{j^c|j}}{{p}_{j|i}{p}_{j^c|j}+{p}_{j^c|i}{p}_{j|j^c}}\right)\frac{1}{n-1}\sum_{t=2}^{n}\mathbbm{I}_{Y_{2t}=\{i,j,{j}^c\}}.
\end{align*}
The first term goes to $0$ (follows from \eqref{trans-p-conv} by first invoking continuous mapping theorem to conclude each term inside the bracket converges a.s. to zero from which it also follows that running average converge to 0) and the second term converges to $\log\left(\frac{2{p}_{j|i}{p}_{j^c|j}}{{p}_{j|i}{p}_{j^c|j}+{p}_{j^c|i}{p}_{j|j^c}}\right)p_ip_{j|i}p_{j^c|j}$ (follows from \eqref{ind-conv}) almost surely. Hence,
\begin{equation*}
    \frac{1}{n-1}\sum_{t=2}^{n}\hat{D}_{2t} \to \sum_{i=0}^1\sum_{j=0}^1 \log\left(\frac{2{p}_{j|i}{p}_{j^c|j}}{{p}_{j|i}{p}_{j^c|j}+{p}_{j^c|i}{p}_{j|j^c}}\right) p_ip_{j|i}p_{j^c|j}, \text{ almost surely, as } n \to \infty,
\end{equation*}

which implies, $\frac{\log M_{2n}}{2n}\to r$, almost surely, as $n\to \infty$, where

\begin{align*}
  r=&\frac{1}{2}\sum_{i=0}^1\sum_{j=0}^1 \log\left(\frac{2{p}_{j|i}{p}_{j^c|j}}{{p}_{j|i}{p}_{j^c|j}+{p}_{j^c|i}{p}_{j|j^c}}\right) p_ip_{j|i}p_{j^c|j}\\
  =& \frac{p_{0|1}p_{1|0}}{2(p_{0|1}+p_{1|0})}\log\Bigg(
  \left(\frac{2(1-p_{0|1})}{1-p_{0|1}+p_{1|0}}\right)^{1-p_{0|1}}
  \left(\frac{2p_{0|1}}{1+p_{0|1}-p_{1|0}}\right)^{p_{0|1}}
  \left(\frac{2p_{1|0}}{1-p_{0|1}+p_{1|0}}\right)^{p_{1|0}}\\
  &~~~~~~~~~~~~~~~~~~~~~~~\times\left(\frac{2(1-p_{1|0})}{1+p_{0|1}-p_{1|0}}\right)^{1-p_{1|0}}\Bigg)\\
  =&\frac{p_{0|1}p_{1|0}}{2(p_{0|1}+p_{1|0})}\Bigg[\log4-(1-p_{0|1})\log\left(1+\frac{p_{1|0}}{1-p_{0|1}}\right)-{p_{1|0}}\log\left(1+\frac{1-p_{0|1}}{p_{1|0}}\right)\\
  &~~~~~~~~~~~~~~~~~~-(1-p_{1|0})\log\left(1+\frac{p_{0|1}}{1-p_{1|0}}\right)-{p_{0|1}}\log\left(1+\frac{1-p_{1|0}}{p_{0|1}}\right)\Bigg].\\
\end{align*}

Since $\log$ is a concave function, we can use Jensen's inequality to obtain
$$r \geq \frac{p_{0|1}p_{1|0}}{p_{0|1}+p_{1|0}}[\log4-(1-p_{0|1}+p_{1|0})\log2-(1-p_{1|0}+p_{0|1})\log2]= 0,$$
where the equality holds if and only if $\frac{p_{1|0}}{1-p_{0|1}}=\frac{1-p_{0|1}}{p_{1|0}}$ and $\frac{p_{1|0}}{1-p_{0|1}}=\frac{1-p_{0|1}}{p_{1|0}}$, which is equivalent to $p_{1|0}=p_{1|1}$.
\qed

\subsection{Proof of Theorem \ref{thm:gen}}
Note that, by definition of ${p}_{j|i}$ and $p_{i,j,j^c}$, we have
\begin{equation}
\label{trans-p-conv-2}
  \hat{p}_{j|i}(t)=\frac{{n}_{j|i}(t)}{{n}_{j|i}(t)+{n}_{j^c|i}(t)} =\frac{{n}_{j|i}(t)/t}{{n}_{j|i}(t)/t+{n}_{j^c|i}(t)/t} \to {p}_{j|i}, \text{ almost surely, as } t \to \infty, \text{ for } i,j\in\{0,1\},
\end{equation}
\begin{equation}
\label{ind-conv-2}
    \frac{1}{n-1}\sum_{t=2}^{n}\mathbbm{I}_{Y_{2t}=\{i,j,{j}^c\}}\to p_{i,j,j^c}, \text{ almost surely, as } n \to \infty, \text{ for } i,j\in\{0,1\}.
\end{equation}
Hence, by the same argument, as shown in the proof of Theorem 2.1, we have
$\frac{\log M_n}{2n}\to r'$, almost surely, as $n\to \infty$, where
\begin{align*}
  2r'
  =&\sum_{i=0}^1\sum_{j=0}^1 \log\left(\frac{2{p}_{j|i}{p}_{j^c|j}}{{p}_{j|i}{p}_{j^c|j}+{p}_{j^c|i}{p}_{j|j^c}}\right) p_{i,j,j^c}\\
  =&p_{1,1,0} \log
  \left(\frac{2(1-p_{0|1})}{1-p_{0|1}+p_{1|0}}\right)
  +p_{0,1,0}\log\left(\frac{2p_{0|1}}{1+p_{0|1}-p_{1|0}}\right)
 +p_{1,0,1}\log\left(\frac{2p_{1|0}}{1-p_{0|1}+p_{1|0}}\right)\\
  &+p_{0,0,1}\log\left(\frac{2(1-p_{1|0})}{1+p_{0|1}-p_{1|0}}\right)\\
  =&(p_{1,1,0}+p_{0,1,0})\left[\log2-\frac{p_{1,1,0}}{p_{1,1,0}+p_{0,1,0}}\log\left(1+\frac{p_{1|0}}{1-p_{0|1}}\right)-\frac{p_{1,1,0}}{p_{1,1,0}+p_{0,1,0}}\log\left(1+\frac{1-p_{0|1}}{p_{1|0}}\right)\right]\\
  &+(p_{1,0,1}+p_{0,0,1})\left[\log2-\frac{p_{1,0,1}}{p_{1,0,1}+p_{0,0,1}}\log\left(1+\frac{p_{0|1}}{1-p_{1|0}}\right)-\frac{p_{0,0,1}}{p_{1,0,1}+p_{0,0,1}}\log\left(1+\frac{1-p_{1|0}}{p_{0|1}}\right)\right].\\
  \geq& (p_{1,1,0}+p_{0,1,0})\left[\log2-\log\left(1+\frac{ap_{0|1}^2+(1-a)(1-p_{0|1})^2}{p_{0|1}(1-p_{0|1})}\right)\right]\\
  &+(p_{1,0,1}+p_{0,0,1})\left[\log2-\log\left(1+\frac{bp_{1|0}^2+(1-b)(1-p_{1|0})^2}{p_{1|0}(1-p_{1|0})}\right)\right]~~\text{(by using Jensen's inequality)}\\
  =& (p_{1,1,0}+p_{0,1,0})\left[\log2-\log\left(2+\frac{(2p_{0|1}-1)(a+p_{0|1}-1)}{p_{0|1}(1-p_{0|1})}\right)\right]\\
  &+(p_{1,0,1}+p_{0,0,1})\left[\log2-\log\left(2+\frac{(2p_{1|0}-1)(b+p_{1|0}-1)}{p_{1|0}(1-p_{1|0})}\right)\right]\\
  \geq & 0, \text{ when } (2p_{0|1}-1)(a+p_{0|1}-1)\geq 0\text{ and } (2p_{1|0}-1)(b+p_{1|0}-1)\geq 0.
\end{align*}
Note that equality holds in the first inequality (which follows from Jensen's inequality) if and only if $\frac{p_{1|0}}{1-p_{0|1}}=\frac{1-p_{0|1}}{p_{1|0}}$ and $\frac{p_{1|0}}{1-p_{0|1}}=\frac{1-p_{0|1}}{p_{1|0}}$, which is equivalent to $p_{1|0}=p_{1|1}$.
Therefore, $r'>0$ if $(2p_{0|1}-1)(a+p_{0|1}-1)\geq 0\text{ and } (2p_{1|0}-1)(b+p_{1|0}-1)\geq 0$ and $p_{1|0}\neq p_{1|1}$.

\subsection{Proof of Theorem \ref{thm:ar-rate}}
Define, $C_{2t}=\log S_{2t}$, and $\hat{C}_{2t}=\log \hat{S}_{2t}$. So, we have $\log(W_{2n})=\sum_{t=2}^n\hat{C}_{2t}$.

Note that $C_{2t}=\log\left(\frac{2f(X_{2t-2},X_{2t-1},X_{2t})}{f(X_{2t-2},X_{2t-1},X_{2t})+f(X_{2t-2},X_{2t},X_{2t-1})}\right)$ is a continuous  function of $ X_{2t},X_{2t-1},X_{2t-2}$. Since, under the alternative, $\{X_t\}_t$ is a stationary AR(1) process, it is an ergodic process and so is the process $\{C_{2t}\}_t$. Now, using ergodic theorem, we can directly say that
\begin{equation}
   \frac{1}{n-1}\sum_{t=2}^nC_{2t}\to \mathbb{E}(C_4) = \mathbb{E}\log(S_4), \text{ almost surely as } n\to\infty.
\end{equation}
Also, using ergodic theorem, it can be shown that $\hat{a}_{t-1} =\sum_{i=2}^{t-1}X_iX_{i-1}/\sum_{i=1}^{t-2}X_i^2 \stackrel{a.s}{\rightarrow} a$ and $\hat{\sigma}^2_{t-1}=\frac{1}{t-2}\sum_{i=2}^{t-1}(X_{i}-\hat{a}_{t-1}X_{i-1})^2 \stackrel{a.s}{\rightarrow} \sigma^2$, as $t\to \infty$, which implies $\hat{C}_{2t} - C_{2t}\stackrel{a.s}{\rightarrow} 0$, as $t\to \infty$.
Then, 
\begin{equation}
    \frac{1}{n-1}\sum_{t=2}^n\hat{C}_{2t} - \frac{1}{n-1}\sum_{t=2}^nC_{2t}\stackrel{a.s}{\rightarrow} 0.
\end{equation}
Hence, 
\begin{equation}
\label{eq:conv}
    \frac{\log W_{2n}}{2n} =\frac{n-1}{2n} \times \frac{1}{n-1}\sum_{t=2}^n\hat{C}_{2t} \to \frac{1}{2}\mathbb{E}\log(S_4) \text{ almost surely, as } n \to \infty,
\end{equation}
\begin{align}
\label{eq:ineq1-2}
   \nonumber r^*=\frac{1}{2}\mathbb{E}\log(S_4)&=\frac{1}{2}\log2-\frac{1}{2}\mathbb{E}\left(\log\left(1+\frac{f(X_2,X_4,X_3)}{f(X_2,X_3,X_4)}\right)\right)\\
   &\geq \frac{1}{2}\log2-\frac{1}{2}\log \left(1+\mathbb{E}\left(\frac{f(X_2,X_4,X_3)}{f(X_2,X_3,X_4)}\right)\right)
\end{align}
The last step follows from Jensen's inequality and equality holds if and only if $\log\left(1+\frac{f(X_2,X_4,X_3)}{f(X_2,X_3,X_4)}\right)$ is a linear function in $X_2,X_3,X_4$, which is equivalent to $a=0$.

It can be easily verified that $\mathbb{E}(\log f(X_2,X_3,X_4))=-1-\log(2\pi) \text{ and }$
\begin{equation}
\label{eq:exp}
    \mathbb{E}(\log f(X_2,X_4,X_3))=-1-\frac{a^2}{1+a}-\log(2\pi)\leq \mathbb{E}(\log f(X_2,X_3,X_4)),
\end{equation}
which implies 
\begin{align*}
  \log\left(\mathbb{E}\left(\frac{f(X_2,X_4,X_3)}{f(X_2,X_3,X_4)}\right)\right)&\leq \mathbb{E}\left(\log\left(\frac{f(X_2,X_4,X_3)}{f(X_2,X_3,X_4)}\right)\right)\\
  &=\mathbb{E}(\log f(X_2,X_4,X_3))-\mathbb{E}(\log f(X_2,X_3,X_4))\leq 0,  
\end{align*}

i.e, $\mathbb{E}\left(\frac{f(X_2,X_4,X_3)}{f(X_2,X_3,X_4)}\right)\leq 1$, where equality holds iff $a=0$ (follows from Equation \eqref{eq:exp}).

Hence, from Equation \eqref{eq:ineq1-2}, $r^*\geq 0$ and equality holds if and only if $a=0$. \qed

\subsection{Proof of Theorem \ref{thm:cont-gen}}

Note that in the proof of Theorem 2.5, we only required
$\{X_t\}_t$ to an ergodic process, in order to conclude Equation \eqref{eq:conv}. Hence, given $\{X_t\}_t$ to be an ergodic process, we have
\begin{equation}
    \frac{\log W_{2n}}{2n} \to \frac{1}{2} \mathbb{E}\log(S_4) \text{ almost surely, as } n \to \infty,
\end{equation}
\begin{equation}
\label{eq:ineq1}
    r^*=\frac{1}{2}\mathbb{E}\log(S_4)=-\frac{1}{2}\mathbb{E}\log\left(\frac{1}{S_4}\right)\geq -\log \mathbb{E}\left(\frac{1}{S_4}\right)
\end{equation}
which follows from Jensen's inequality and equality holds if and only if $\log\left(1+\frac{f(X_2,X_4,X_3)}{f(X_2,X_3,X_4)}\right)$ is a linear function in $X_2,X_3,X_4$, which is equivalent to $a=0$.
Now, given that $\mathbb{E}\left(\frac{1}{S_4}\right)\leq 1,$ we have $r^*\geq 0$. Moreover, $r^*> 0$, if $a\neq0$ and $\mathbb{E}\left(\frac{1}{S_4}\right)\leq 1$. \qed

\subsection{Proof of Theorem \ref{thm:mc-triple}}

For $t\geq 2$, let us define 
$Y_{t}=(X_{3t},X_{3t+1},X_{3t+2},X_{3t+3})$,
$\hat{D}^*_{t}=\log \hat{B}^*_{t}$. So, we can write $\log(M^*_{n})=\sum_{t=2}^n\hat{D}^*_{t}$.

Now, by using ergodic theorem, we can show that the MLE of the transition probability from $i$ th state to $j$ th state, $\hat{p}_{j|i}$ is strongly consistent for ${p}_{j|i}$, i.e,
\begin{equation}
\label{a-trans-p-conv}
    \hat{p}_{j|i}(t)\to {p}_{j|i}, \text{ almost surely, as } t \to \infty, \text{ for } i,j\in\{0,1\}.
\end{equation}
And since $X_1,X_2,\cdots$ is a Markov chain of first order, $Y_1,Y_2,\cdots$ is also a Markov chain of first order. So, again by ergodic theorem,
\begin{equation}
\label{a-ind-conv}
    \frac{1}{n-1}\sum_{t=1}^{n-1}\mathbbm{I}_{Y_{t}=\{i,j,k,l\}}\to p_ip_{j|i}p_{k|j}p_{l|k}, \text{ almost surely, as } n \to \infty, \text{ for } i,j,k,l\in\{0,1\},
\end{equation}
where $p_i=\frac{p_{i|i^c}}{p_{i|i^c}+p_{i^c|i}}.$ is the stationary probability of $i$th state. Now,
\begin{align*}
    &\frac{1}{n-1}\sum_{t=2}^{n}\hat{D}^*_{t}\\
    &=\frac{1}{n-1}\sum_{t=2}^{n} \sum_{i=0}^1\sum_{\substack{j,k,l=0\\(j,k,l)\neq\\(0,0,0),(1,1,1)}}^1 \log\left(\frac{3\hat p_{j|i}(t)\hat p_{k|j}(t)\hat p_{l|k}(t)}{\sum_{\pi \in \Pi(j,k,l)} \hat p_{\pi(j)|i}(t)\hat p_{\pi(k)|\pi(j)}(t)\hat p_{\pi(l)|\pi(k)}(t)}\right)\mathbbm{I}_{Y_{t}=\{i,j,k,l\}}\\
    &= \sum_{\substack{j,k,l=0\\(j,k,l)\neq\\(0,0,0),(1,1,1)}}^1 \frac{1}{n-1}\sum_{t=2}^{n} \Bigg[\log\left(\frac{3\hat p_{j|i}(t)\hat p_{k|j}(t)\hat p_{l|k}(t)}{\sum_{\pi \in \Pi(j,k,l)} \hat p_{\pi(j)|i}(t)\hat p_{\pi(k)|\pi(j)}(t)\hat p_{\pi(l)|\pi(k)}(t)}\right)\\
    &~~~~~ -\log\left(\frac{3p_{j|i}p_{k|j}p_{l|k}}{\sum_{\pi \in \Pi(j,k,l)} p_{\pi(j)|i}p_{\pi(k)|\pi(j)}p_{\pi(l)|\pi(k)}}\right)\Bigg]\mathbbm{I}_{Y_{t}=\{i,j,k,l\}}\\
    &~~~~~+\sum_{\substack{j,k,l=0\\(j,k,l)\neq\\(0,0,0),(1,1,1)}}^1 \log\left(\frac{3p_{j|i}p_{k|j}p_{l|k}}{\sum_{\pi \in \Pi(j,k,l)} p_{\pi(j)|i}p_{\pi(k)|\pi(j)}p_{\pi(l)|\pi(k)}}\right)\frac{1}{n-1}\sum_{t=2}^{n}\mathbbm{I}_{Y_{t}=\{i,j,k,l\}}.
\end{align*}
The first term goes to $0$ (follows from \eqref{a-trans-p-conv} by first invoking continuous mapping theorem to conclude each term inside the bracket converges a.s. to zero from which it also follows that running average converge to 0) and the second term converges to $\log\left(\frac{3p_{j|i}p_{k|j}p_{l|k}}{\sum_{\pi \in \Pi(j,k,l)} p_{\pi(j)|i}p_{\pi(k)|\pi(j)}p_{\pi(l)|\pi(k)}}\right)p_ip_{j|i}p_{k|j}p_{l|k}$ (follows from \eqref{a-ind-conv}) almost surely. Hence,
\begin{equation*}
    \frac{1}{n-1}\sum_{t=2}^{n}\hat{D}^*_{2t} \to \sum_{i=0}^1\sum_{\substack{j,k,l=0\\(j,k,l)\neq\\(0,0,0),(1,1,1)}}^1\log\left(\frac{3p_{j|i}p_{k|j}p_{l|k}}{\sum_{\pi \in \Pi(j,k,l)} p_{\pi(j)|i}p_{\pi(k)|\pi(j)}p_{\pi(l)|\pi(k)}}\right) p_ip_{j|i}p_{k|j}p_{l|k}., 
\end{equation*}
$\text{ almost surely, as } n \to \infty,$
which implies, $\frac{\log M^*_{n}}{3n}\to \Tilde{r}'$, almost surely, as $n\to \infty$, where

\begin{align*}
  \Tilde{r}'=&\frac{1}{3}\sum_{i=0}^1\sum_{\substack{j,k,l=0\\(j,k,l)\neq\\(0,0,0),(1,1,1)}}^1\log\left(\frac{3p_{j|i}p_{k|j}p_{l|k}}{\sum_{\pi \in \Pi(j,k,l)} p_{\pi(j)|i}p_{\pi(k)|\pi(j)}p_{\pi(l)|\pi(k)}}\right) p_ip_{j|i}p_{k|j}p_{l|k}.\\
  =& -\frac{1}{3}\sum_{i=0}^1 p_ic_i\sum_{\substack{j,k,l=0\\(j,k,l)\neq\\(0,0,0),(1,1,1)}}^1 \frac{p_{j|i}p_{k|j}p_{l|k}}{c_i}\log\left(\frac{\sum_{\pi \in \Pi(j,k,l)} p_{\pi(j)|i}p_{\pi(k)|\pi(j)}p_{\pi(l)|\pi(k)}}{3p_{j|i}p_{k|j}p_{l|k}}\right).\\
  \geq & \frac{1}{3}\sum_{i=0}^1p_ic_i \log(3c_i) -\frac{1}{3}\sum_{i=0}^1 p_ic_i \log \Bigg(\sum_{\substack{j,k,l=0\\(j,k,l)\neq\\(0,0,0),(1,1,1)}}^1 \sum_{\pi \in \Pi(j,k,l)} p_{\pi(j)|i}p_{\pi(k)|\pi(j)}p_{\pi(l)|\pi(k)}\Bigg)=0\\
\end{align*}
Since $\log$ is a concave function, we have use Jensen's inequality here and equality holds iff for fixed $i\in \{0,1\}$, $\frac{\sum_{\pi \in \Pi(j,k,l)} p_{\pi(j)|i}p_{\pi(k)|\pi(j)}p_{\pi(l)|\pi(k)}}{3p_{j|i}p_{k|j}p_{l|k}}$ takes the same value for all $j,k,l\in \{0,1\}, (j,k,l)\neq (0,0,0),(1,1,1)$, which implies $p_{1|1}=p_{0|1}$.
\qed

\subsection{Proof of Theorem \ref{thm:ar-triple}}

Define, $C^*_{t}=\log S^*_{t}$, and $\hat{C}^*_{t}=\log \hat{S}^*_{t}$. So, we have $\log(W^*_{n})=\sum_{t=2}^n\hat{C}^*_{t}$.

Note that $C^*_{t}=\log\left(\frac{6g(X_{3t},X_{3t+1},X_{3t+2},X_{3t+3})}{\sum_{i=1}^6g(X_{3t},X^{i}_{t,1},X^{i}_{t,2},X^{i}_{t,3})}\right)$ is a continuous  function of $ X_{3t},X_{3t+1},X_{3t+2}$ and $X_{3t+3}$. Since, under the alternative, $\{X_t\}_t$ is a stationary AR(1) process, it is an ergodic process and so is the process $\{C^*_{t}\}_t$. Now, using ergodic theorem, we can directly say that
\begin{equation}
   \frac{1}{n-1}\sum_{t=2}^nC^*_{t}\to \mathbb{E}(C^*_2) = \mathbb{E}\log(S^*_2), \text{ almost surely as } n\to\infty.
\end{equation}
Also, using ergodic theorem, it can be shown that $\hat{a}_{t-1} =\sum_{i=2}^{t-1}X_iX_{i-1}/\sum_{i=1}^{t-2}X_i^2 \stackrel{a.s}{\rightarrow} a$ and $\hat{\sigma}^2_{t-1}=\frac{1}{t-2}\sum_{i=2}^{t-1}(X_{i}-\hat{a}_{t-1}X_{i-1})^2 \stackrel{a.s}{\rightarrow} \sigma^2$, as $t\to \infty$, which implies $\hat{C}^*_{t} - C^*_{t}\stackrel{a.s}{\rightarrow} 0$, as $t\to \infty$.
Then, 
\begin{equation}
    \frac{1}{n-1}\sum_{t=2}^n\hat{C}^*_{t} - \frac{1}{n-1}\sum_{t=2}^nC^*_{t}\stackrel{a.s}{\rightarrow} 0.
\end{equation}
Hence, 
\begin{equation}
\label{a-eq:conv}
    \frac{\log W^*_{n}}{3n} =\frac{n-1}{3n} \times \frac{1}{n-1}\sum_{t=2}^n\hat{C}^*_{t} \to \frac{1}{3}\mathbb{E}\log(S^*_2) \text{ almost surely, as } n \to \infty,
\end{equation}
\begin{align}
\label{a-eq:ineq1-2}
   \nonumber \Tilde{r}^*=\frac{1}{3}\mathbb{E}\log(S^*_2)&=\frac{1}{3}\log6-\frac{1}{3}\mathbb{E}\left(\log\left(\frac{\sum_{i=1}^6g(X_{3},X^{i}_1,X^{i}_2,X^{i}_3)}{g(X_{3},X_{4},X_{5},X_{6})}\right)\right)\\
   &\geq \frac{1}{3}\log6-\frac{1}{3}\log \left(\sum_{i=1}^6\mathbb{E}\left(\frac{g(X_{3},X^{i}_1,X^{i}_2,X^{i}_3)}{g(X_3,X_4,X_5,X_6)}\right)\right)
\end{align}
The last step follows from Jensen's inequality and equality holds if and only if $\log\left(\frac{\sum_{i=1}^6g(X_{3},X^{i}_1,X^{i}_2,X^{i}_3)}{g(X_{3},X_{4},X_{5},X_{6})}\right)$ is a linear function in $X_3, X_4, X_5, X_6$, which is equivalent to $a=0$.

It can be easily verified that 
\begin{equation}
\label{a-eq:exp}
    \mathbb{E}(\log g(X_{3},X^{i}_1,X^{i}_2,X^{i}_3))\leq\mathbb{E}(\log g(X_3,X_4,X_5,X_6)),
\end{equation}
which implies 
\begin{align*}
  \log\left(\mathbb{E}\left(\frac{g(X_{3},X^{i}_1,X^{i}_2,X^{i}_3)}{g(X_3,X_4,X_5,X_6)}\right)\right)&\leq \mathbb{E}\left(\log\left(\frac{g(X_{3},X^{i}_1,X^{i}_2,X^{i}_3)}{g(X_3,X_4,X_5,X_6)}\right)\right)\\
  &=\mathbb{E}(\log g(X_{3},X^{i}_1,X^{i}_2,X^{i}_3))-\mathbb{E}(\log g(X_3,X_4,X_5,X_6))\leq 0,  
\end{align*}

i.e, $\mathbb{E}\left(\frac{g(X_{3},X^{i}_1,X^{i}_2,X^{i}_3)}{g(X_3,X_4,X_5,X_6)}\right)\leq 1$, where equality holds iff $a=0$ (follows from Equation \eqref{a-eq:exp}).

Hence, from Equation \eqref{a-eq:ineq1-2}, $\Tilde{r}^*\geq 0$ and equality holds if and only if $a=0$. \qed


\bibliography{ref}

\end{document}